\documentclass{aa} 
\usepackage{graphicx}

\usepackage{siunitx}
\usepackage{txfonts}
\usepackage{comment}
\usepackage{float}
\usepackage[dvipsnames]{xcolor}
\usepackage{subfig}
\usepackage[utf8]{inputenc}
\usepackage[T1]{fontenc}
\usepackage{amsfonts}
\usepackage{pdflscape}
\usepackage{longtable}
\usepackage{xcolor}
\usepackage{lscape}
\usepackage{nicefrac}
\usepackage[colorlinks=true,allcolors=blue]{hyperref}
\usepackage{orcidlink}
\usepackage{dblfloatfix}

\usepackage{multirow} 

\usepackage{arydshln} 
\usepackage{caption}

\usepackage{amsfonts}
\usepackage{natbib}

\usepackage{nicefrac}

\usepackage[export]{adjustbox}

\newcommand{\hers}{{\it  Herschel}}

\newcommand{\spitz}{{\it   Spitzer}}
\newcommand{\iso}{{\it  ISO}}

\newcommand{\akari}{{\it  AKARI}}

\newcommand{\sofia}{{\it SOFIA}}
\newcommand{\lmcplus}{{LMC$^{+}$}}
\newcommand{\kms}{{\,km\,s$^{-1}$}}
   \newcommand{\mic}{$\mu$m}

  \newcommand{\zsol}{Z$_\odot$}

  \newcommand{\xdor}{30$\;$Doradus}

  \newcommand{\co}{CO$\,$(1-0)}
   \newcommand{\hmol}{H$_2$}
  
  \newcommand{\halpha}{H$\,${$\alpha$}}

  \newcommand{\hii}{H$\,${\sc ii}}

  \newcommand{\cii}{[C$\,${\sc ii}]}
  
  \newcommand{\oiii}{[O$\,${\sc iii}]}
 \newcommand{\oi}{[O$\,${\sc i}]}

  \newcommand{\ciiline}{\cii$\lambda 158$\mic}

  \newcommand{\oiiilineup}{\oiii$\lambda 88$\mic}

\newcommand{\ciico}{L$_\mathrm{[C\,{\sc II}]}$/L$_\mathrm{CO(1-0)}$}
\newcommand{\ciifir}{L$_\mathrm{[C\,{\sc II}]}$/L$_\mathrm{FIR}$}

\newcommand{\oiiicii}{L$_\mathrm{[O\,{\sc III}]}$/L$_\mathrm{[C\,{\sc II}]}$}
\newcommand{\ciitir}{L$_\mathrm{[C\,{\sc II}]}$/L$_\mathrm{TIR}$}

\newcommand{\cotir}{L$_\mathrm{[C\,{\sc O}]}$/L$_\mathrm{TIR}$}
\newcommand{\lcii}{L$_\mathrm{[C\,{\sc II}]}$}

\newcommand{\lco}{L$_\mathrm{CO}$}
\newcommand{\lfir}{L$_\mathrm{FIR}$}

\newcommand{\ltir}{L$_\mathrm{TIR}$}

\newcommand{\tir}{L$_\mathrm{TIR}$}
\newcommand{\fir}{L$_\mathrm{FIR}$}
\newcommand{\intensity}{W m$^{-2}$ sr$^{-1}$}

\graphicspath{{./}{figures/}}

\begin{document} 
\title{\lmcplus: Large-scale mapping of \cii\ and \oiii\ in the LMC molecular ridge}
\subtitle{I. Dataset and line ratio analyses}
\author{C. Fischer\inst{1,2}\thanks{Corresponding author: Christian Fischer \newline \email{fischer@iram.fr}}\orcidlink{0000-0003-2649-3707}, S.~C.~Madden\inst{3}\orcidlink{0000-0003-3229-2899}, A.~Krabbe\inst{1,4}\orcidlink{0000-0002-8522-7006}, F.~L.~Polles\inst{3,5}\orcidlink{0000-0003-0347-3201}, D.~Fadda\inst{6}\orcidlink{0000-0002-3698-7076}, E.~ Tarantino\inst{6}\orcidlink{0000-0003-1356-1096}, F.~Galliano\inst{3}\orcidlink{0000-0002-4414-3367}, C.-H.~R.~Chen\inst{7}\orcidlink{0000-0002-3925-9365}, N.~Abel\inst{8}\orcidlink{0000-0003-1791-723X}, Á.~Beck\inst{1}\orcidlink{0000-0002-1373-1377}, L. Belloir\inst{3}\orcidlink{0009-0005-7593-9789}, F.~Bigiel\inst{9}\orcidlink{0000-0003-0166-9745}, A.~Bolatto\inst{10}\orcidlink{0000-0002-5480-5686}, M.~Chevance\inst{11,12}\orcidlink{0000-0002-5635-5180}, S.~ Colditz\inst{1}\orcidlink{0000-0002-5613-1953}, N.~Fischer\inst{1}\orcidlink{0009-0003-0641-8993},  A.~Green\inst{13}\orcidlink{0000-0002-8432-3362}, A.~Hughes\inst{14}\orcidlink{0000-0002-9181-1161}, R.~Indebetouw\inst{15}\orcidlink{0000-0002-4663-6827}, C.~Iserlohe\inst{1,4}\orcidlink{0000-0003-4223-7439}, M.~Kaźmierczak-Barthel\inst{1}\orcidlink{0000-0002-6024-0060}, R.~Klein\inst{16}\orcidlink{0000-0002-7187-9126}, A.~Lambert-Huyghe\inst{3}, V.~Lebouteiller\inst{17}\orcidlink{0000-0002-7716-6223}, E.~Mikheeva\inst{18}\orcidlink{0009-0008-4342-2664}, A.~Poglitsch\inst{3}\orcidlink{0000-0002-6414-9408},  L.~Ramambason\inst{11}\orcidlink{0000-0002-9190-9986}, W.~Reach\inst{19}\orcidlink{0000-0001-8362-4094}, M.~Rubio\inst{20}\orcidlink{0000-0002-5307-5941},  W.~Vacca\inst{21}\orcidlink{0000-0002-9123-0068}, T.~Wong\inst{13}\orcidlink{0000-0002-7759-0585}, H.~Zinnecker\inst{22}\orcidlink{0000-0003-0504-3539}}

\institute{$^{1}$ Deutsches SOFIA Institut, University of Stuttgart, Pfaffenwaldring 29, 70569, Stuttgart, Germany\\
$^{2}$ IRAM - Institut de Radioastronomie Millimétrique, 300 rue de la Piscine, 38406 Saint-Martin d'Hères, France\\
$^{3}$ Université Paris Cité, Université Paris-Saclay, CEA, CNRS, AIM, 91191, Gif-sur-Yvette, France\\
$^{4}$ Institute of Space Systems - SOFIA Data Center, University of Stuttgart, Pfaffenwaldring 29, 70569, Stuttgart, Germany\\
$^{5}$ SOFIA Science Center, USRA, NASA Ames Research Center, M.S. N232-12 Moffett Field, CA, 94035, USA \\
$^{6}$ Space Telescope Science Institute, 3700 San Martin Dr., Baltimore, MD 21218, USA\\
$^{7}$ Max-Planck-Institut f\"ur Radioastronomie, Auf dem
  H\"ugel 69, 53121 Bonn, Germany \\
$^{8}$ University of Cincinnati, Clermont College, 4200 Clermont College Drive, Batavia, OH, 45103, USA\\
$^{9}$ Argelander-Institut für Astronomie, Auf dem Hügel 71, 53121 Bonn, Germany\\
$^{10}$ Department of Astronomy, University of Maryland, College Park, MD 20742, USA\\
$^{11}$ Universit\"{a}t Heidelberg, Zentrum f. Astronomie, Institut f. Theoretische Astrophysik, A.-Ueberle-Str 2, 69120 Heidelberg, Germany \\
$^{12}$ Cosmic Origins Of Life (COOL) Research DAO, \href{https://coolresearch.io}{coolresearch.io} \\
$^{13}$ Astronomy Department, University of Illinois, Urbana, IL 61801, USA\\
$^{14}$ CNRS, IRAP, 9 Av. du Colonel Roche, BP 44346, F-31028 Toulouse cedex 4, France \\
$^{15}$ Department of Astronomy, University of Virginia, P.O. Box 3818, Charlottesville, VA 22903, USA \\
$^{16}$ Lockheed Martin Solar \& Astrophysics Laboratory, Palo Alto, CA 94304, USA \\
$^{17}$ Université Paris-Saclay, Université Paris-Cité, CEA, CNRS, AIM, 91191, Gif-sur-Yvette, France\\
$^{18}$ Astro Space Center of P.N. Lebedev Physics Institute (ACS LPI), Moscow, 117997, Russia\\
$^{19}$ Space Science Institute, 4765 Walnut Street, Suite 205, Boulder, CO 80301, USA\\
$^{20}$ Departamento de Astronomía, Universidad de Chile, Casilla 36-D, Santiago, Chile\\
$^{21}$ NSF’s NOIRLab, 950 N. Cherry Avenue, Tucson, AZ 85719, USA\\
$^{22}$ Universidad Autónoma de Chile, Nucleo Astroquimica y Astrofisica, Avda Pedro de Valdivia 425, Providencia, Santiago de Chile, Chile
}
\date{Received June 05, 2025; accepted September 01, 2025} 
\titlerunning{\lmcplus\ }
\authorrunning{Fischer et al., 2025}

\abstract
   {The fundamental process of star formation in galaxies involves the intricate interplay between the fueling of star formation via molecular gas and the feedback from recently formed massive stars that can, in turn, hinder the conversion of gas into stars. This process, by which galaxies evolve, is also closely connected to the intrinsic properties of the interstellar medium (ISM), such as structure, density, pressure, metallicity, etc.}
   {To study the role that different molecular and atomic phases of the ISM play in star formation, and to characterize their physical conditions, we zoom into our nearest neighboring galaxy, the Large Magellanic Cloud (LMC; 50 kpc), the most convenient laboratory to study the effects of the lower metal abundance on the properties of the ISM. The LMC offers a view of the ISM and star formation conditions in a low metallicity ($Z\sim0.5$\,\zsol) environment similar to, in that regard, the epoch of the peak of star formation in the earlier universe ($z\sim1.5$). Following up on studies carried out at galactic scales in low-Z galaxies, we present an unprecedentedly detailed analysis of a well-known star-forming regions (SFRs) at a spatial resolution of a few pc.}
   {We mapped a $610pc \times 260pc$ region in the LMC molecular ridge in \ciiline\ and the \oiiilineup\ using the FIFI-LS instrument on the SOFIA telescope. We compare the data with the distribution of the CO (2-1) emission from ALMA, the modeled total infrared luminosity as well as \spitz/MIPS 24\mic\ continuum and H$\alpha$.} 
   { We present new large maps of \cii\ and \oiii\ and perform a first comparison with CO (2-1) line and \ltir\ emission. We also provide a detailed description of the observing strategy with \sofia/FIFI-LS and the data reduction process.}
   {We find that \cii\ and \oiii\ emission is associated with the SFRs in the molecular ridge, but also extends throughout the mapped region, not obviously associated with ongoing star formation. The CO emission is clumpier than the \cii\ emission and we find plentiful \cii\ present where there is little CO emission, possibly holding important implications for CO-dark gas. We find a clear trend of the \ciitir\ ratio decreasing with increasing \ltir\ in the full range. This suggests a strong link between the “\cii-deficit“ and the local physical conditions instead of global properties.}
\maketitle

\section{Introduction}

\begin{figure*}[!t]
\includegraphics[width = 0.502\textwidth]{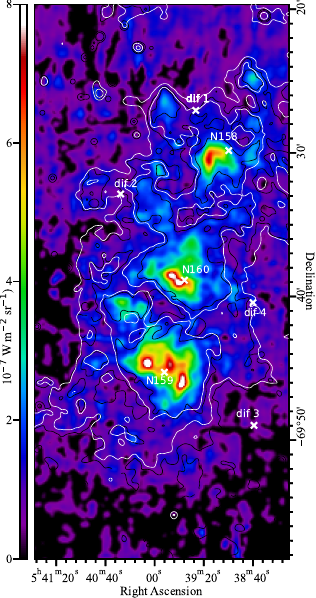}
\includegraphics[width = 0.478 \textwidth]{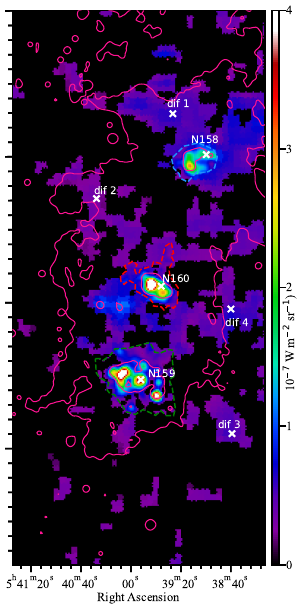}
\caption{Intensity of the far-infrared fine-structure lines \ciiline\ (left) and \oiiilineup\ (right) in \intensity, observed by \sofia/FIFI-LS. The intensities are the result of spectral Gaussian fits and only those fits with a S/N $\geq$ 3  are shown. Pixels with lower S/N are displayed as 0 (in black).  The white contours are from \ltir\ (300 and 600 L$_{sol}$/pc$^2$) shown in Fig. \ref{fig:LTIR}. The thin black contours shown in the \cii\ map are 8\mic\ from \spitz/IRAC (4 and 8 MJy/Sr). The red contours in the \oiii\ map are \spitz/MIPS 24\mic\ continuum (5 and 150 MJy/sr). The \cii\ map is shown in the native resolution with a beam FWHM of 15.3\arcsec. The \oiii\ data in the three bright star formation regions marked by blue, green and red dashed lines) was smoothed to the same resolution (15.3\arcsec) to increase the S/N per beam. Outside of the regions the data was further re-binned to 24\arcsec\ and fitted within a 48\arcsec\ beam to further increase the S/N in the diffuse regions. The 7 gray crosses, in identical positions in the \cii\ and \oiii\ maps, mark the positions of the spectra shown in Appendix \ref{spectra}.}
\label{fig:fifi_maps}
\end{figure*}

The necessary fuel to feed star formation is considered to be molecular gas, as evidenced, for example, by the Kennicutt-Schmidt relationship \citep{kennicutt98}. As molecular hydrogen cannot be observed directly, the carbon monoxide (CO) J=1–0 transition has traditionally been the convenient tracer of total molecular gas and is thus a calibrator of star formation activity in the local universe as well as at high redshifts.
However this leads to puzzling results in low-metallicity (low-Z) galaxies. There the relation of molecular gas traced by \co\ to the efficiency with which these galaxies form stars is perplexing. This is mainly due to the fact that the search for \co\ in low-Z galaxies is a veritable challenge \citep[e.g.,][]{cormier14, madden20}, making the conversion of observed \co\ to molecular hydrogen (\hmol)  in low-Z galaxies fraught with uncertainty \citep[e.g.,][]{Schruba12, bolatto13,cormier14, madden20, ramambason24}.\\
\begin{table*}
\caption{Observational parameters of \sofia/FIFI-LS observations; the radial velocity was used to center the arrays spectrally and the beam size in pc was calculated based on a distance of 50 kpc; the integration time $T_{int}$ is per FIFI-LS spatial pixel. 
\label{table:observations}} 
\centering 
\begin{tabular}{c c c c c c c c c}
\hline \hline 
Channel & Line & Transition & \multicolumn{1}{c}{$\lambda_\mathrm{rest}$} & rad. velocity&\multicolumn{1}{c}{$FWHM_{\rm{Beam}}$} & \multicolumn{1}{c}{$FWHM_{\rm{Beam}}$} & $T_{\rm{int}}$ & R  \\
~ & & & \multicolumn{1}{c}{(\mic)} & (km/s) & ($\arcsec$)& (pc)& \multicolumn{1}{c}{(s)} & \\ 
\hline
~blue & [OIII] &  $^3$P$_{1}$--$^3$P$_{0}$ & 88.356 &  220 & 8.6 & 2.1 & 10 & 620 \\
~red& [CII] & $^2$P$_{3/2}$--$^2$P$_{1/2}$ & 157.741 & 220 & 15.3 & 3.7 & 40 & 1200 \\

\hline 
\hline 
\end{tabular}
\end{table*}
Reduced metal abundance in galaxies has important consequences on the structure and physical properties of the Interstellar Medium (ISM), especially in the vicinity of intense radiation fields, arising from the atmospheres of metal-poor stars and carried by winds of the most massive stars. The combination of hard radiation fields and the low dust abundance leads to widespread ionization and significant photodissociation of CO molecules in the large H$^0$-\hmol\ shells of low extinction. As a result, CO becomes increasingly difficult to detect in low-Z dwarf galaxies, while self-shielded \hmol\ may exist outside the CO-emitting core, making it "CO-dark". These conditions allow the ultraviolett (UV) photons to carve out a rather porous ISM structure \citep[e.g.,][]{chevance16,Grishunin2024}, facilitating feedback processes by further channeling the photons afar. In studies using the Dwarf Galaxy Survey \citep[DGS;][]{madden13} the porosity of the ISM increases as Z decreases \citep{cormier19, Ramambason2022}, and that this property is also sensitive to resolution, with smaller regions ($\sim$200pc) being more porous to ionizing photons than larger ones \citep{Polles2019}.

While CO is difficult to detect in low-Z galaxies \citep[e.g.,][]{madden13,cormier14}, the dominant cooling line of the ISM in galaxies, \ciiline, remains relatively bright compared to higher-Z galaxies \citep{stacey91,madden20}. This is illustrated in the low-Z galaxy, IC10, suggesting that more molecular gas was present than that traced by CO alone \citep{madden97}. Subsequent surveys of \cii\ in local, low-Z star-forming galaxies have confirmed the elevated \ciico\ ratios \citep{cormier15, Hunter01}. The \ciiline\ line has been suggested as a tracer of CO-dark \hmol\ gas, which has become an increasingly important component in the low-Z ISM \citep[e.g.,][]{madden20, ramambason24},  helping to quantify the total \hmol\ in galaxies.\\
Dissecting the structure of the ISM in low-Z, star-forming galaxies by targeting distinct gas phases with appropriate diagnostic tracers, allows us to reveal critical details of the transition from atomic to molecular gas. This approach can provide a deeper understanding of the formation and distribution of \hmol\ and a better handle on quantifying and characterizing the total \hmol\ reservoir in low-Z galaxies.\\ 
The Large Magellanic Cloud (LMC) is our closest low-Z galaxy (50 kpc, e.g. \cite{schaefer08}) and offers a view of the ISM and star formation conditions in a low-Z ($Z\sim0.5$\,\zsol) environment similar to, in that regard, the epoch of the peak of star formation in the earlier universe ($z\sim1.5$) \citep[e.g.,][]{chruslinska19}. It serves as an ideal laboratory for examining the effect of massive star formation on the surrounding ISM, zooming into the well-known regions like \xdor\ \citep{Pellegrini2011,chevance20,okada2019_2,Grishunin2024}, N11 \citep{lebouteiller12} and N159 \citep{Okada15,Lee2016,Fukui2015,Fukui2019,Tokuda2019}. High fractions of CO-dark molecular gas were found in 36 beams throughout the LMC with the \hers\ Heterodyne Instrument for the Far-Infrared (HIFI) by \cite{Pineda2017} and \cite{Pellegrini2012} found a high fraction of optically thin H II regions showing that UV photons can often escape them and penetrate the surrounding ISM.  In \xdor\ the transition from the ionized nebular gas around the massive R136 star cluster to the neutral photodissociation region (PDR) is prominently traced by the extensive \cii\ emission. This emission highlights UV-illuminated clouds extending well beyond the central cluster, leaving fragments of CO clouds which may be embedded in a substantial reservoir of "hidden", CO-dark gas \citep[e.g.][]{chevance20,wong22}. However, while \xdor\ is indeed a spectacular region, it is a unique and very extreme case illuminated by a massive super star cluster, R136, and not representative of the diversity of star formation conditions in the LMC or other galaxies.\\
To explore this issue further, we have mapped an extensive region of the LMC, covering 260pc $\times$ 610pc in the \ciiline\ and the \oiiilineup\ emission lines with the Stratospheric Observatory for Infrared Astronomy (\sofia) Legacy Program, \lmcplus. These new maps enable us to examine not only the effects of massive to moderate star formation but also the far-reaching effects of star formation on the ISM throughout the range of environmental conditions in the molecular ridge, just south of \xdor. This study aims to understand the role of the gas component that is traced by \ciiline, its association with molecular clouds, and its link to star formation, especially in  metal-deficient environments.\\ 
Importantly, \cii\ is one of the most popular species targeted in observations of high-Z galaxies and serves as a critical tracer of star formation across cosmic time \citep[e.g.,][]{delooze14,lagache18,Schaerer2020}. Understanding its behavior in low-Z environments like the LMC is essential for interpreting observations of galaxies in the early universe \citep[see e.g.,][]{Wolfire2022,HerreraCamus2025,Zanella2018,Lefevre2020,Bouwens2022}.\\
\cii\ was first observed in the LMC  with far infrared heterodyne receiver of the Kuiper Airborne Observatory by \citet{Boreiko91} in multiple single beams, including some pointings toward the molecular ridge star-forming regions (SFRs), N158, N159, and N160 (marked in Fig. \ref{fig:fifi_maps}).  The molecular ridge hosts the largest accumulation of CO in the LMC \citep{Cohen1988, Fukui2008, Kutner1997, Mizuno2001}. The northern portion of this $\sim 2$kpc feature hosts the three bright SFRs while the southern portion is quiescent with little high-mass star formation \citep[see][]{Finn2021, Indebetouw2008}. Measurements of both \oiii\ and \cii\ have been made with the Infrared Space Observatory (\iso) Long Wavelength Spectrometer (LWS) \citep{Vermeij02a} and the \akari\ Far-Infrared Surveyor / Fourier Transform Spectrometer \citep{Kawada11}, both with limited angular resolution and very limited spatial coverage. Two 80\arcsec\ beams have been analyzed from \iso/LWS centered on N159 and N160 in \oiii\ and \cii. Most of the (10'$\times$1.5') stripes  with a Full Width at Half Maximum (FWHM) of 40\arcsec\ from \akari\ are more or less centered around \xdor, where \oiii\ detections are found about 150 pc away from the R136 cluster, suggesting a porous ISM structure as the gas remains excited by R136 at these distances.  Four stripes of observations are available, as well, with \akari\, targeting the bright SFRs of the molecular ridge, with some detections reaching beyond the star-forming sites. Around the \xdor\ region, a correlation is found between \oiii\ and H$\alpha$ emission \citep [from the SHASSA survey; ][]{GaustadtSHASSA2001}. However, across the entire dataset, this correlation becomes less clear for high \oiii\ intensities. The difference in ionization degree of the gases is suspected as a potential reason for that. In the \akari\ fields of the molecular ridge, which include some areas beyond the bright SFRs, \oiii\ and H$\alpha$ fluxes are relatively low and do not show any correlation. \cite{Kawada11} also see a global correlation between \oiii\ and the continuum at 88 \mic\ which is mainly driven by data points close to \xdor\ which offer a wide range of both line flux and continuum values. Again there is no correlation in their data from the molecular ridge. The fact that \akari\ detected  \oiii\ off the bright star formation regions in the ridge provides a strong motivation for a larger-scale map of \oiii\ in the molecular ridge to probe the porosity and ionization structure of the ISM in diverse environments in the LMC.\\
\cii\ has been observed at high spatial and spectral resolution toward the bright SFRs N158, N160 and N159 with limited mapping using \sofia/upGREAT (\citealp{Okada15}; \citealp{Okada19}). They find that \cii\ typically has a wider line profile than CO and that in those bright regions, 30$\%$ of the \cii\ emission is not reproducible by the CO-defined line profile, with a lower fraction toward molecular clouds. While this might indicate shielded clumps along the line of sight, it may also indicate a possible small contribution from CO-dark gas. On the other hand, it has been suggested that up to $\sim$90$\%$ of the molecular gas in the \xdor\ region originates from CO-dark gas \citep{chevance20}. Thus, it is compelling to study the diverse molecular ridge  on a large scale, covering very different SFRs as well as much more extended regions beyond the limited SFRs.

\section{Observations and data reduction}\label{sec:obs}
\subsection{\sofia\ FIFI-LS}\label{obs:fifi}

Observations of the \cii\ and the \oiii\ lines were carried out with the Field-Imaging Far-Infrared Line Spectrometer (FIFI-LS) \citep{Fischer18,Colditz18} on board \sofia\ \citep{Erickson93,Young12}. FIFI-LS is an integral-field imaging spectrometer that provides simultaneous observations in two channels: the blue channel covering 51-125 \mic\ and the red channel covering 115-203 \mic.  Each channel consists of an overlapping 5$\times$5 pixel footprint on the sky, where the pixel size is $6\arcsec$ and $12\arcsec$ for the blue and red channels, respectively.  We refer to each of these spatial pixels as \arcsec spaxels\arcsec. The optics within FIFI-LS rearranges the 25 spaxels into a pseudo-slit, and the light impinging on each spaxel is then dispersed spectrally (using a grating) over 16 pixels. This generates an integral-field data cube for each observation.  The spectral resolution is wavelength dependent ranging from $R = \lambda$/$\Delta\lambda \simeq$ 500 to 2000.

\begin{figure}[h]
\includegraphics[width = 0.49\textwidth]{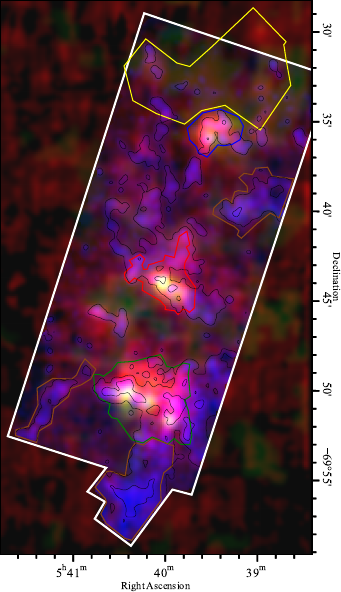}
\caption{RGB map of \cii\ (r), \oiii\ (g) and $\rm ^{12}CO(2-1)$ (b)  from Tarantino et al. (in preparation) and Chen et al. (in preparation), with the map of the $\rm ^{12}CO(2-1)$ outlined in white. Contours are shown for $\rm ^{12}CO(2-1)$ line flux levels of $5*10^{11}$ and $5*10^{10} W m^{-2} sr^{-1}$. Regions for analyses in Sect.~\ref{sec:dis} are outlined, corresponding respectively to the SFRs, N159(green), N160(red), N158(blue), an extended, less active region in \oiii\ (yellow) and three CO-bright filaments (brown). They are also referred to in Fig.~ \ref{fig:LTIR}, and in the plots in Fig.~\ref{fig:ratio_diagrams} and Fig.~\ref{fig:oiii_24}.}
\label{fig:rgb}
\end{figure}

\subsubsection{Observations}
\sofia/FIFI-LS observations were taken in March 2022 as part of the Cycle 9 \sofia\ Legacy Program, \lmcplus, $09_{-}0036$ (PIs S. Madden, A. Krabbe$)$. \sofia\ was on deployment to the southern sky flying out of Santiago de Chile. Of the awarded 20h wall-clock time for the legacy project (50h in total requested) 15h were observed during 6 flights. While the metallicity of the Small Magellanic Cloud is even lower than that of the LMC, the factor of 2 to 4 times lower line fluxes in the SMC \citep{cormier15} would not have permitted a large map detecting \cii\ in the lower brightness ISM conditions.\\
Due to the large size of the mapped region ($\sim$1°$\times$0.5°), observations were performed without chopping in On-The-Fly (OTF) mapping mode for maximum efficiency and best spatial sampling with short integration times per field. In this mode, the telescope scans across the mapping area at a constant speed. The speed ($6\arcsec/s$) was chosen to avoid distortion of the image during the shortest continuous integration, which is 0.125s for FIFI-LS \citep{Fischer18}. The movement of $0.75\arcsec$ during this integration was less than $10$\% of the spatial FWHM at both observed wavelengths. The fine sampling provided by the scan enables mapping with the best possible spatial resolution \citep[see][]{fadda23} with the spatially over-sized FIFI-LS pixels. With a scan length of 30s, $180\arcsec$ were covered on the sky in one scan. With the field of view size of $30\arcsec\times30\arcsec$ in the blue channel, 12 scans were performed to fill the $3'\times3'$ tiles with 2$\times$6 scans carried out in the perpendicular direction to ensure even sampling. The 5$\times$5 spaxel array was tilted by 11.3° to spread out the spaxels evenly, perpendicular to the direction of the scan. At the beginning and the end of each scan, part of the $180\arcsec$ scan was not seen by all of the spaxels. Full integration time there is reached by the overlap of the field of views with the scans from the neighboring tiles, creating an interlock.\\
With the $60\arcsec\times60\arcsec$ footprint of the red array, this mode produced additional overlap of the scans, resulting in a map with four times deeper integration, improved flat-fielding, and a uniformly deep integrated map. The efficiency (on-source integration time vs. wall clock time) was 51$\%$, well above the efficiencies achieved with pointed observations \citep{fischer16}. For background subtraction, a field with minimal expected \cii\ emission\footnote{from a simulation} as well as minimal 250\mic\ continuum emission\footnote{from \hers/SPIRE} was identified in proximity to the mapping region, ensuring an offset smaller than 30$\arcmin$ to all fields in the map. This allowed the telescope to reach the offset position within its limited movement range due to the aircraft's orientation. Observational details for both lines are listed in Table \ref{table:observations}. A more detailed description of the OTF observing mode with FIFI-LS is provided in \cite{Fischer25}.\\
The 3\arcmin$\times$3\arcmin\ interlocking tiles were the building blocks of each sub-map. The central region of the map with the three bright SFRs N158, N159, and N160, was covered in a single rectangular map with 14$\times$6 tiles for a total extension of 42\arcmin$\times$18\arcmin. Integrations were taken tile by tile in rows from the northern edge close to \xdor\ and progressing southward. Originally, 7 flights were planned to complete all 84 tiles, with potential time for expansion further south. However, due to one lost flight, 76.25 tiles ($91$\%) were completed. The relatively high completion rate was achieved through flight re-planning to maximize LMC leg times.

\subsubsection{Data reduction}
The data was reduced with the \sofia/FIFI-LS pipeline \citep{Vacca2020}, which includes all the necessary calibrations and flat-field corrections \citep{fadda23}. We produced cubes with 6, 12 and 24 spatial pixel size for the analysis presented in this paper. Since OTF mapping mode was only offered for FIFI-LS on \sofia\ since Cycle 9 and had never been used on a map of this large size, some modifications to the last \sofia\ published version of the pipeline\footnote{\url{https://github.com/SOFIA-USRA/sofia_redux} } have been applied to achieve better background subtraction, noise propagation, and telluric correction.\\
We modified the noise calculation to better capture all noise contributions in OTF mode, i.e. non linear drifts of the telluric emission. While these are captured in pointed FIFI-LS observations by averaging $\sim$50 data points \citep{fischer16}  with constant signal, this is not done for OTF where every data point on a scan has a different astronomical signal. Since the overall signal is background dominated, we can assume that any non linear behavior of the signal on the scan is dominated by the background and not the astronomical emission. We introduced a new step in the data reduction procedure ("scan noise definition") to quantify this for each scan and combined it with the errors for each data point. More details can be found in \cite{Fischer25}.\\
Due to the large size of the map, nod-offsets of up to $\sim$30' were required. Depending on the rotation of field of the \sofia\ telescope, this can lead to significant changes in telescope elevation, which in turn affected the airmass. At the altitudes \sofia\ operates (37000ft - 45000ft), the raw far-infrared signal is dominated by sky emission. As a result, the gradient in airmass during nodding for background subtraction can introduce significant signal, complicating the background subtraction process. This can also add noise to the baseline in the overlapping, interlocking region of the tiles. The signal offset due to imperfect background subtraction varies over time, generating  artifacts when tiles, taken at different times or even on different flights, overlap. This effect can be amplified by telluric correction since this signal contributions are not transmitted through the atmosphere.\\
Two different approaches to remove artifacts from those airmass differences were applied to the \cii\ and \oiii\ data due to the different telluric features in the spectral ranges of the two lines. They are both described in full detail in \cite{Fischer25}. For \cii, a telluric scaling technique was used where an emission model with contributions from the atmosphere, as well as telescope and instrument background, is fitted to the signal in each of the 25 spaxels in the off-nod. If there are significant telluric features in the spectral range, this two-component model allows the signal in each of the 400 pixels to be split up into the elevation-dependent sky emission and the background that is independent from the telescope elevation. The sky contribution, and with it the overall off-nod signal, were then scaled to the elevation of the on-nod. Telluric scaling was possible for \cii\ but not for \oiii, since there are no significant spectral features in the range used for the \oiii\ line. Thus, the telluric contribution to the \oiii\ signal cannot be quantified. Still, the spectra in the \oiii\ cube display underlying telluric artifacts. It has proven to be most advantageous to remove them during the fitting of the emission line described below.\\
For the telluric correction we used water vapor values obtained with the method described in \citet{fischerPWV} and \citet{iserlohePWV} in the pipeline. With the water vapor overburden, telescope elevation and flight altitude for each scan, a transmission profile is calculated with the ATRAN model \citep{Lord92} for the observed spectral range. In the pipeline it is then spectrally smoothed to the instrument's resolution and applied to each ramp at the observed wavelength in a background-subtracted nod pair. Using the smoothed transmission profiles does not distort the line profiles here since the transmission is spectrally flat in the relevant range in both channels. 
However, since we cannot remove baseline artifacts with telluric scaling for the \oiii\ map, we modified the telluric correction procedure in the pipeline to avoid further distortion of the baseline. For the \oiii\ map of the molecular ridge, the transmission of the emission line only depends on atmospheric conditions. It remains largely unaffected by variations in local velocity and line width. For each nod cycle, we determined the transmission at the wavelength of the \oiii\ line\footnote{using the rad. velocity from table \ref{table:observations}} and applied this correction to the entire spectrum.  This preserves the original baseline shape as measured. For the \oiii\ data, this approach enables telluric correction of the line flux based on the actual atmospheric conditions during the scan, while also allowing for the fitting of any residual telluric baseline.
\begin{table*}
\caption{Fit parameters used for both lines of \sofia/FIFI-LS observations \label{table:fit}} 
\centering 
\begin{tabular}{c c c c c c c}
\hline \hline 
Channel & Line & Transition & min. wavelength & max. wavelength & line position range  & line width range\\
~ & & & \multicolumn{1}{c}{(\mic)} & \multicolumn{1}{c}{(\mic)}& \multicolumn{1}{c}{(\mic)} & (km/s)\\ 
\hline
~blue & [OIII] & $^3$P$_{1}$--$^3$P$_{0}$ & 88.13-88.29 & 88.57-88.77 &  88.37-88.44 & 450-600 \\
~red& [CII] & $^2$P$_{3/2}$--$^2$P$_{1/2}$ & 157.66-157.74 & 157.93-158.10 &  157.80-157.93 & 200-333 \\
\hline 
\hline 
\end{tabular}
\end{table*}

\subsubsection{\cii\ and \oiii\ line fitting} 
\label{line_fitting}
The spectrally unresolved lines were fitted in circular apertures around each spatial pixel with Gaussian profiles together with the baseline using a SciPy curve\_fit function\footnote{\url{https://docs.scipy.org/doc/scipy/reference/generated/scipy.optimize.curve_fit.html}}. The line widths and center positions were constrained based on the spectral resolution of FIFI-LS (with margins) and the mapped region in the LMC (by multiple iterations of the fit). To avoid edge effects due to the re-binning, portions of the spectra near the spectral cube edges were cut. All fit parameters are listed in Table \ref{table:fit}. For \cii, where telluric artifacts were removed using telluric scaling in the pipeline processing, a zero-order polynomial was fitted to the baseline. For \oiii, a residual telluric emission model was fitted to the baseline. The model includes a zero-order polynomial and atmospheric transmission from ATRAN, which can be scaled positively or negatively to account for residual telluric emission caused by nodding to different elevations with the telescope. More details on this "telluric fitting" can be found in \cite{Fischer25}. Examples of line profiles and fits are shown in Appendix \ref{spectra}.\\  
The integrated line flux maps of \cii\ and \oiii\ are shown in Fig. \ref{fig:fifi_maps}. The map of \cii\ is shown in the native telescope resolution, with a FWHM of 15.3\arcsec used as aperture diameter in the line fitting. The \oiii\ data was also fitted with this aperture to improve the signal-to-noise ratio (S/N) per beam. They are shown together with the CO (2-1) line flux in Fig. \ref{fig:rgb}. For the maps in Figs. \ref{fig:fifi_maps} and \ref{fig:rgb}, the data was spatially re-binned to 6\arcsec\ and fitted within 15.3\arcsec\ diameter apertures around each spatial pixel for both \cii\ and \oiii.  It was increased to 24\arcsec\ with a 48\arcsec\ diameter apertures outside the bright star formation regions for \oiii. For all ratio plots the binning was increased to 12\arcsec\ for \cii\ and \oiii\ in the bright SFRs. The FIFI-LS data was smoothed to 18\arcsec resolution for ratio plots involving \ltir.

\subsection{Cross calibration}

A few bright regions of the LMC were observed with the \hers\ Photodetector Array Camera and Spectrometer (PACS)~\citep{cormier15}. This gave us the opportunity to cross calibrate the fluxes obtained with \sofia/FIFI-LS. We were able to define 20 circular regions (12 for \cii\ and 8 for \oiii) to perform such cross-correlation (see Table~\ref{table:PACS_cross}). For each region we defined circular apertures with a 18\arcsec\ radius. Such apertures were sufficiently larger than the PSF so that the measured fluxes were not affected by the spread of point-like emissions.
Since the PACS data were obtained in "unchopped mode", we reprocessed them using the "transient-correction" pipeline (see \cite{fadda16} and \cite{sutter22b} for recent updates). In fact, as pointed out in \cite{fadda23}, the PACS data in the \hers\ Science Archive can deviate significantly since the flux calibration was based on the internal calibrators rather than on the more stable telescope background as is the case of normal chop-nod observations. Sudden flux changes when switching between source and internal calibrators can induce a major, unpredictable transient in the signal, varying the response of the detectors up to 90\%.
The fluxes were evaluated with pseudo-Voigt curves which provide the best fitting and recover most of the flux in the wings of the spectra for the higher S/N in the PACS spectra (see~\cite{fadda23}). 
With an average deviation below 5$\%$ (3$\%$ for \cii\ and 4.6$\%$ for \oiii), the agreement between FIFI-LS  and PACS data is excellent and well within the estimated flux uncertainties of PACS and FIFI-LS \citep{PACS_Quick, fadda23}.
  
\subsection{Ancillary data}\label{obs:other}

\begin{figure}
\includegraphics[width = 0.4\textwidth]{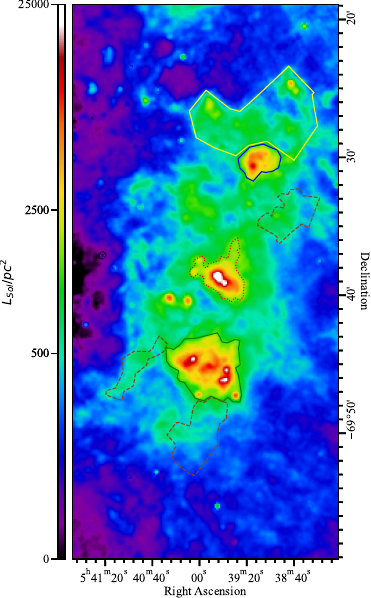}
\caption{Map of TIR power in the molecular ridge. Regions for the analyses in Chapter \ref{sec:dis} are outlined, corresponding, respectively, to the SFRs, N159(green), N160(red dotted), N158(blue), an extended, less active region in \oiii\ (yellow), and three CO-bright filaments (brown dashed). They are also shown in Fig. \ref{fig:rgb}, and used in the plots in Figs. \ref{fig:ratio_diagrams}, \ref{fig:CII_LTIR}, and  \ref{fig:oiii_24}. }
\label{fig:LTIR}
\end{figure}

\subsubsection{ALMA CO (2-1)}
To map the molecular ridge in CO(2-1) as a tracer of molecular gas it was observed with the Atacama Large Millimeter/Submillimeter Array (ALMA) Atacama Compact Array (ACA), also known as the Morita array. These observations were part of Cycle 7, under project code 2018.A.00061.S (PI: Bolatto, A). The map covers a region of 10\arcmin\ $\times$ 26\arcmin\ (150 pc $\times$ 380 pc) across the northern tip of the molecular ridge in the LMC, overlapping with three \hii\ regions: N158, N160, and N159 and the majority of the \sofia\ observations. The map is comprised of sixteen individual 10' $\times$ 2.3' tiles. One frequency setting was configured to cover the $\rm ^{12}CO(2-1)$ line at 230.538 GHz with a bandwidth of 125 MHz and a channel width of 61 kHz. The second frequency setting covered both $\rm ^{13}CO(2-1)$ at 220.399 GHz and $\rm C^{18}O(2-1)$ at 219.560 GHz with bandwidths of 62.5 MHz and a channel width of 61 kHz. We also included a broader continuum frequency set-up centered at 232.6 GHz with a bandwidth of 2 GHz. These observations include total power ALMA data corresponding to the full rectangular region covering the interferometric map produced on October 15th, 2019, providing short spacings information.\\
We used version 6.5.1 of the Common Astronomy Software Applications \citep{casa2022} and the standard system calibration to image the molecular ridge data. We imaged each of the sixteen individual tiles that comprise the final map separately and stitched them together to create the final mosaic. Imaging each tile separately provides higher stability and convergence in the de-convolution algorithm \citep[see also][]{Leroy2021}. For the spectral line data, we used the CASA task \texttt {sdint} to image and combine the \textit{ACA} and TP data. The \texttt{sdint} algorithm is a joint de-convolution algorithm that simultaneously images and combines interferometric and single dish data \citep{Rau2019}. We found that the \texttt{sdint} algorithm produces better results than the traditional \texttt {tclean} and \texttt{feather} method. Each tile has a square pixel cell size of 1.2\arcsec, a velocity channel of 0.25 \kms, and a total size of 600 $\times$ 300 pixels. We chose 320 channels which corresponds to the maximum size of the bandwidth for the $\rm ^{13}CO(2-1)$ and $\rm C^{18}O(2-1)$ data. We used a cleaning threshold of 0.3 Jy/beam and a \textit{cyclefactor = 5} in order to make sure the cleaning algorithm does not diverge before it reaches the threshold. We used the mosaic gridder, hogbom deconvolver, and briggs weighting with \textit{robust = 0.5}, which provides the best combination of resolution and signal-to-noise. We masked the data for cleaning through the \texttt {auto-multithresh} and the default parameters. The interferometric data were combined with the total power data using an \textit{sdgain = 3}, which preserves the flux information from the total power data while maintaining the high resolution structure. We convolved the data cubes to a common restoring beam of 7\arcsec\ (1.7 pc). The details of the reduction and analysis of the $\rm ^{12}CO(2-1)$, $\rm ^{13}CO(2-1)$, and $\rm C^{18}O(2-1)$ will be presented in Tarantino et al. in prep.\\
The \textit{ACA} CO(2-1) observations of N159's southern region were obtained as part of the ALMA project 2016.1.00782.S (PI: Chen, R.). The map is comprised of two rectangle regions, 96$\arcsec\times$200$\arcsec +$78$\arcsec\times$50\arcsec (23 pc$\times$49 pc$+$20 pc$\times$12 pc) that covers all bright CO filaments detected in the Atacama Pathfinder Experiment (APEX) CO (2-1) mosaic (under project ID M-092.F-0020). The Band 6 receiver setup for the CO(2-1) line has a bandwidth of 117.2 MHz and a channel width of 141 kHz (or 0.184 km~s$^{-1}$); the receiver setup and detailed reduction and analysis for other lines in Band 6 including $^{13}$CO(2-1), C$^{18}$O(2-1), SiO(5-4), H$_2$CO (3$_{0,3}$-3$_{0,2}$, 3$_{2,2}$-3$_{2,1}$, 3$_{2,1}$-3$_{2,0}$), $^{13}$CS(5-4) will be presented in Chen et al.\ (in prep.). The ACA observations do not include total power ALMA data as the APEX CO(2-1) mosaic is available for the short spacing information. The ACA and APEX CO(2-1) data were combined using the \texttt{feather} task under CASA.

\subsubsection{The total infrared map}

The map of the Total InfraRed (TIR; 3-1000\mic) power in our region was computed from its spatially-resolved Spectral Energy Distribution (SED). The dust parameter maps derived from this SED modeling are presented by Belloir et al. (in preparation) using the \spitz\ Infrared Array Camera (IRAC) and Multiband Imaging Photometer for Spitzer (MIPS) maps at $\lambda$=3.6, 4.5, 5.8, 8.0, 24 and 70 \mic\ from the SAGE project \citep{meixner06} and the \hers/PACS and the Spectral and Photometric Imaging Receiver (SPIRE) maps $\lambda$=100, 160, 250 \mic\ from the HERITAGE project \citep{meixner13}.
The angular resolutions of these maps were homogenized to 18\arcsec\ using the convolution kernels of \citet{aniano11} and re-projected onto the \sofia\ [C$\,$\textsc{ii}] pixel grid. The noise uncertainties of the original maps were propagated through the homogenization process using a bootstrapping method \citep{press07}. At the end of this process a spectral cube (RA, Dec, $\lambda$) and its corresponding noise cube were obtained. The homogenized spectral cube was then fitted using the hierarchical Bayesian code, HerBIE \citep{galliano18a}. The model set-up was identical to that used by \citet{galliano21}: a dust mixture, with the properties of the THEMIS dust model \citep{jones17}, illuminated by a range of starlight intensities, $U$, following a power-law distribution and a stellar continuum, approximated as the Rayleigh-Jeans tail of a black body. HerBIE accounts for the fact that calibration uncertainties are spatially fully correlated and spectrally partially correlated. The hierarchical Bayesian approach of this model provides a regularization of the pixels with poor signal-to-noise ratios. The most relevant inferred parameters, for each pixel, are \citep[see][]{galliano18a,galliano21}:
\begin{itemize}
  \item the total dust mass, $M_{\textnormal{\footnotesize dust}}$;
  \item the fraction of small amorphous carbon grains, carrying the mid-IR aromatic 
    features, $q_{\textnormal{\footnotesize AF}}$;
  \item the mean starlight intensity, $\langle U\rangle$;
  \item the TIR luminosity, $L_{\textnormal{\footnotesize TIR}}$ (Figure \ref{fig:LTIR}).
\end{itemize}

\subsubsection{\spitz\ MIPS and MCELS H$\alpha$}
The \spitz/MIPS 24 \mic\ data for the molecular ridge was available on the IRSA Infrared Science Archive \footnote{https://irsa.ipac.caltech.edu/data/SPITZER/SAGE/}. The H$\alpha$\ data is from the Magellanic Cloud Emission-line Survey (MCELS, \cite{points24}) available at the NOIRLab Astro Data Archive \footnote{https://astroarchive.noirlab.edu/portal/results/collection/DeMCELS/}. They are used as tracers of exposed (H$\alpha$) and embedded (24 \mic) star formation in comparison with the \oiii\ data to investigate the distribution of the dust and gas densities, the role of the excitation sources/radiation field and the porosity.

\section{Results}\label{sec:Results}
\subsection{\cii\ emission}
\cii\ is detected with S/N $\geq$ 3 almost everywhere in the mapped region (Fig. \ref{fig:fifi_maps}). The highest fluxes are observed towards the three bright SFRs, but there is also significant diffuse emission throughout the mapped region, especially compared to \ltir, which drops off steeply beyond the SFRs (Fig. \ref{fig:LTIR}). The good correlation of \cii\ with the 8$\mu m$ continuum observed toward the bright SFRs by \citet{Okada19} is also obvious in our map and holds as well for the extended structures shown by the black 8$\mu m$ contours in Fig. \ref{fig:fifi_maps}. The spatial correlation between the \cii\ emission and 8$\mu m$ emission is seen even for the low \cii\ fluxes (values < \num{3e-7} \intensity), except in the northern and western parts of the map, where \cii\ emission is more extended than the 8$\mu m$ emission. Comparison of the \cii\ map to the distribution of CO (2-1), shows that the SFRs are bright in both tracers, as expected (and in \oiii, as well). In the more diffuse regions, well beyond the star-forming sites, however, there is no recognizable correlation spatially between \cii\ and CO (2-1). Fig. \ref{fig:rgb} also shows how diffuse the \cii\ emission compares to the filamentary structure of the CO (2-1) emission.

\subsection{\oiii\ emission}
While \cii\ emission is well detected almost everywhere in the mapped region, \oiii\ is more challenging to detect. This is mainly due to the higher noise level of the \oiii\ map caused by the smaller integration time per beam of the blue channel. To analyze the data, we smoothed the map to the same resolution as \cii\ in the three bright star formation regions (15.3\arcsec). However, to increase the S/N per beam everywhere else in the map, the map was re-binned to 24\arcsec\ with a 48\arcsec\ aperture for the line fits (Fig. \ref{fig:fifi_maps}) there. The resulting map shows that the  \oiii\ emission, which traces ionized gas, is more localized and compact compared to the \cii\ emission. 
The brightest \oiii\ emission is concentrated towards the three major SFRs, dropping off rapidly beyond the star-forming sites. Yet, \oiii\ emission is still detected in large areas of the map even outside of the bright regions. This is consistent with the detections of \oiii\ by \akari\ in the molecular ridge region \cite{Kawada11}. Where \oiii\ is detected, it is typically brighter than \cii\ (see Fig. \ref{fig:ratio_diagrams}). The bright \oiii\ emission spatially correlates with the 24 $\mu m$ emission and with the \ltir\, which are usually attributed to dust heating by young, massive stars and serve as a good tracers of ongoing dust-obscured star formation \citep[e.g., ][]{Gratier2012,Corbelli2017,Kim2021}. Notably, there is an extended region directly north of N158 (marked in yellow in Figs. \ref{fig:rgb} and \ref{fig:LTIR}), which has extended clear detections of \oiii, although not particularly bright in \cii\ or \ltir.

\section{Discussion} 
\label{sec:dis}
\begin{figure*}
\centering
\includegraphics[width=0.488\textwidth]{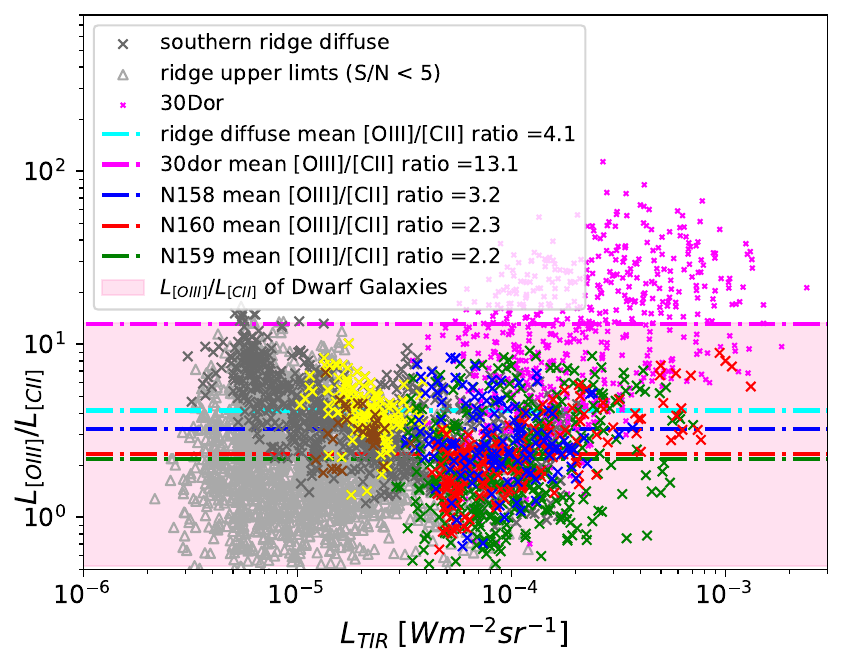}
\includegraphics[width=0.49\textwidth]{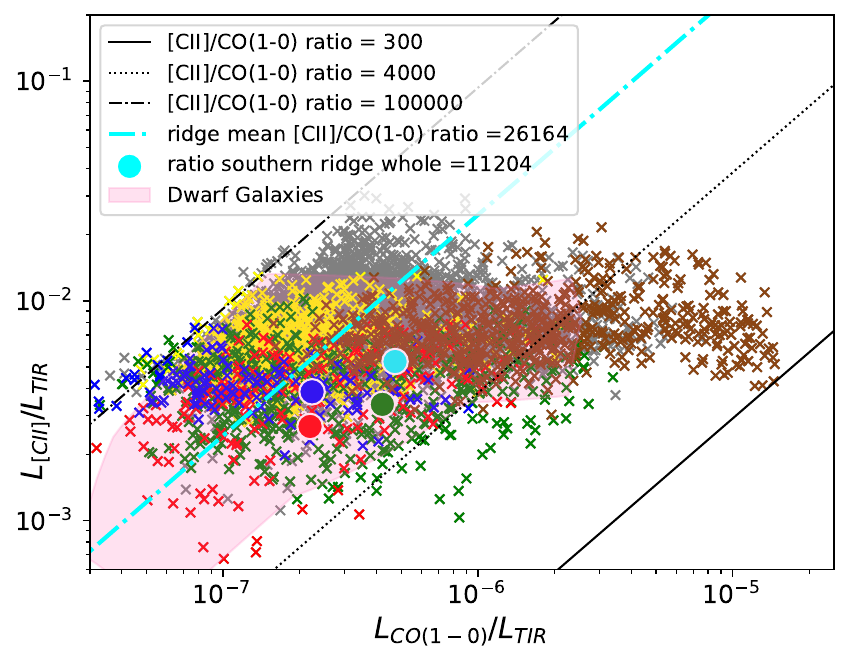}
\caption{
Line ratio plots for the molecular ridge. Yellow, blue, red, green and brown colors refer to the regions outlined in Figs. \ref{fig:rgb} and \ref{fig:LTIR}. Gray points represent the pixels in the ridge outside of these identified regions. 
{\it Left}: \oiiicii\ versus \tir\ in 12\arcsec\ pixels for the three bright star formation regions and in 24\arcsec\ pixels for the rest of the molecular ridge, where pixels with S/N below 5 are shown as upper limits.  \xdor\ data points from \cite{chevance20} have been added with 12\arcsec\ binning. The horizontal lines show the mean \oiiicii\ ratio for the regions. {\it Right}:
\ciitir\ versus \cotir\ in 12\arcsec\ pixels. CO (1-0) was estimated from CO (2-1) assuming a ratio of CO (2-1)/CO (1-0) of 0.6 (e.g. \citealt{denBrok2021}; \citealt{Maeda2022}). The average ratio (no fit) of \ciico\ from the molecular ridge is indicated with the cyan line. The large circles represent the ratios based on the fluxes for the whole regions, in cyan for the whole molecular ridge and in matching colors for the three star formation regions. The area filled in red represents the data range for dwarf galaxies from \cite{madden20} using a \tir\ to \fir\ ratio of 0.6. 
}
\label{fig:ratio_diagrams}
\end{figure*}

In this section, we investigate the relationship between \cii\ and \oiii, and other tracers to characterize the physical properties of the region. Due to the ionization potential of carbon being close to but smaller than that of hydrogen, \cii\ can trace gas from H II regions as well as PDRs. Studies have shown that the ionized phase often contributes negligibly to its emission \citep[e.g.,][]{Pineda2013,Tarantino2021}. In particular, in low-Z environments [CII] originates predominantly from PDRs rather than from H II regions \citep{cormier19, Lebouteiller2019, Ramambason2022}. In general, any \cii\ arising from the ionized gas decreases as the radiation field increases.  Towards \xdor, one of the most massive  SFRs in the LMC, it has been shown that at least 90$\%$ of the \cii\ emission originates in PDRs \citep{chevance16}. 
To aid in the interpretation of the ratio plots, we highlight several regions, which are shown in the three-color image in Fig. ~\ref{fig:rgb} and the \ltir\ image of Fig. \ref{fig:LTIR}. The three bright SFRs, N158 (blue), N160 (red), and N159 (green), were delineated by hand based on the \ltir\ emission (Fig. \ref{fig:LTIR}) at a level of $\sim$1000 L$_{sol}$/$pc^2$. Some regions based on bright CO (2-1) emission relative to \ltir\ are outlined in brown and a region of relatively bright diffuse \oiii\ emission, located north of N158, is outlined in yellow. These regions are noted in these colors in the plots of Figs.  \ref{fig:CII_LTIR}, \ref{fig:ratio_diagrams} and \ref{fig:oiii_24}.

\subsection{\cii\ compared to TIR}
\begin{figure}
\includegraphics[width = 0.49\textwidth]{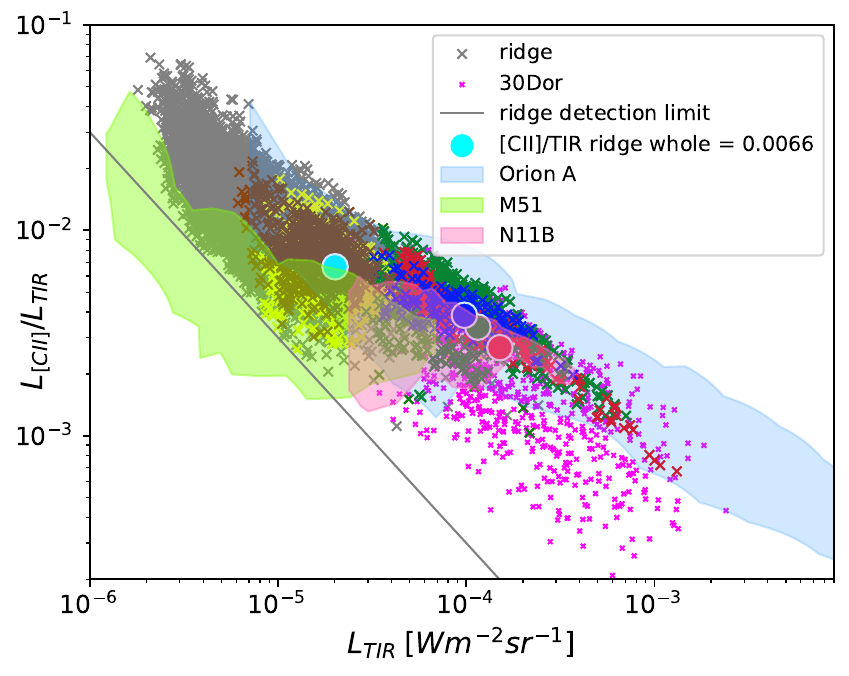}
\caption{\ciitir\ versus \ltir\ in the molecular ridge in 12\arcsec × 12\arcsec\ pixels. Yellow, blue, red, green and brown colors refer to the regions outlined in Figs. \ref{fig:rgb} and \ref{fig:LTIR}. Gray points represent the pixels in the ridge outside of these regions. Data for \xdor\ from \cite{chevance20} is added. The gray line represents the average minimum detectable \cii\ line flux for S/N = 3. The large circles represent the ratio values for the whole ridge (cyan) and the three bright SFRs (N158: blue; N160: red; N159:green), relative to the mean \ltir. The areas with filled colors represent observations from: Orion A \citep{Pabst21}; M51 \citep{Pineda18}; N11B \citep{lebouteiller12}).}
\label{fig:CII_LTIR}
\end{figure}

In conditions of thermal equilibrium within PDRs, the heating due to the radiation from young stars should be balanced by the cooling through line emission. Dust absorbs UV radiation emitted by stars, and primarily cools by re-emitting in the infrared. Small grains and  polycyclic aromatic hydrocarbons (PAHs), when heated by stellar radiation, release electrons that heat the surrounding gas, via the process of photoelectric heating - the dominant heating mechanism in PDRs (see e.g. review by \cite{Wolfire2022}). The fraction of energy in the dust heating that is transferred to the gas heating is measured with the photoelectric heating efficiency, which is the ratio of gas heating to dust heating. Since \cii\ emission is the main cooling channel in diffuse PDRs, and \ltir\ accounts for all contributions from the dust emission, the \ciitir\ ratio has been used as a tracer of the photoelectric heating efficiency \cite[e.g.,][]{Diaz13, cormier15}. Several studies (e.g.\citealt{Malhotra97}; \citealt{Gracia11}; \citealt{Diaz17}; \citealt{HerreraCamus2018}) have shown that the \ciitir\ ratio decreases with the increase of \ltir\ for emission integrated over full galaxies. This "\cii-deficit" affects the reliability of \cii\ as a tracer of star formation. \cite{HerreraCamus2018} investigated the "\cii-deficit" in a sample of unresolved galaxies using toy models as well as Cloudy simulations. Their findings suggest that the observed decrease of \ciifir\ can be explained by an increase in the ionization parameter with \lfir. The models indicate that as the ionization parameter increases, the far-infrared emission increases, but the photoelectric heating efficiency declines, leading to a reduction in \ciifir. Several recent studies using higher spatial resolution data from \hers\ and \sofia\, resolving parts of galaxies or fully mapping them (e.g.\citealt{Lutz16}; \citealt{smith17}; \citealt{Pineda18}; \citealt{Bigiel20}; \citealt{sutter22a}) have shown that the \cii\ deficit  is not driven by global properties of galaxies but rather by local physical properties. These studies find that the \lcii\ to \ltir\ ratio continuously declines with increasing \ltir. A similar trend was also shown within the Milky Way with a large-scale \cii\ map in Orion A by \cite{Pabst21}, confirming this behavior.

Figure \ref{fig:CII_LTIR} shows the \ciitir\ ratio as a function of \ltir\ (Belloir et al. in preparation) for the molecular ridge as well as \xdor\ (\citealt{chevance20}). The \ciitir\ ratio for the entire molecular ridge is about 0.7\% and the three bright SFRs with around 0.3\% fall well within the range for entire galaxies in the DGS sample (0.04\%-0.8\%) from \cite{cormier15}. The values throughout the molecular ridge span about three orders of magnitude in \ltir\ and about two orders of magnitude in the \ciitir\ ratio, with a clear trend of decreasing \ciitir\ as \ltir\ increases. The ridge regions extended beyond the main SFRs (gray points in Fig. ~\ref{fig:CII_LTIR}) are located in the upper left of the plot, exhibiting the highest \ciitir\ ratios and lowest \ltir\ values, followed by the regions with bright CO (2-1) emission (brown) and regions with relatively bright diffuse \oiii\ emission (yellow) located outside the main SFRs, north of N158.  Next in the sequence are the data of the SFRs of the ridge, which in parts are bright in \ltir\ and overlap with the \xdor\ data points, that show very low \ciitir\ with high \ltir.\\
Throughout the molecular ridge, \ciitir\ maintains a clear trend, with less scatter than in \xdor, varying by no more than $\sim$ 0.5 orders of magnitude from a straight trend. At lower \ltir, the trend is potentially affected by the noise limit of the \cii\ map, but remains clear, as \cii\ is detected in the large majority of the mapping area. The three brightest SFRs exhibit a pronounced decline in \ciitir\ in their brightest regions, compared to the overall trend. In these dense regions, \oi\ line emission likely contributes significantly to the cooling(as seen in N11 in \cite{lebouteiller12}).  However, the clear and continuous drop in \lcii\ relative to \ltir\ throughout the dynamic range strongly suggests a correlated decrease of the photoelectric heating efficiency with increasing \ltir.\\
For the bright SFRs, \ltir\ spans a similar range to that in the \xdor\ map. The \ciitir\ ratio falls into the same range as in \xdor. However, for most of the data points of \xdor, the \ciitir\ ratio is lower than in any part of the molecular ridge. The ionizing sources in \xdor, which are more energetic than those of N158, N159, and N160, lead to increased ionization of carbon ions, reducing \cii\ emission relative to \tir \citep{chevance16}.\\
At lower spatial resolution, \cite{Israel96} observed enhanced \cii\ emission relative to infrared emission around N159, attributing it to deep UV penetration due to low dust-to-gas ratio. Similarly, \cite{madden97} encountered elevated \ciitir\ in some more diffuse regions of the low-Z galaxy IC10, suggesting that the high ratio indicates the presence of low UV fields and deep UV penetration caused by the low dust-to-gas ratio. This also is thought to result in a large fraction of CO-dark molecular gas in those regions where  \cii\ presumably is tracing significant amounts of molecular hydrogen. The trend also agrees with the findings in M33 by \cite{Kramer20}, where a different \ltir\ definition was used, preventing precise comparison.\\
Figure~\ref{fig:CII_LTIR} also displays the range of values from N11B \citep{lebouteiller12} in the LMC, as well as large regions mapped in Orion A \citep{Pabst21}, and M51 \citep{Pineda18}.  The spread of values altogether show good agreement, following the sequence the ridge observations,  across different environments and scales, revealing a consistent trend: \ciitir\ increases at lower \ltir\ values, with a tight correlation. This suggests a strong link between the "\cii-deficit" and local physical conditions. A more detailed analysis of the \ciitir\ ratio and possible origins of the high values found in the diffuse parts of the ridge is presented in Belloir et al. (in preparation).

\subsection{\oiii\ compared to \cii}
\label{oiii_cii}

The ratio of \oiiicii\ over  \tir\ is shown in Fig.~\ref{fig:ratio_diagrams} left. For comparison, data for \xdor\ from \citet{chevance20} are shown as well. In the diffuse regions, we display pixels with S/N > 5 for both emission lines, while pixels with lower S/N are also included but marked with different symbols. The range of ratios observed closely matches the range of ratios integrated over the low-Z dwarf galaxies from the DGS sample \citep{cormier15}. While \oiii\ line flux is  on average 13 times that of \cii\ in the \xdor\ map, the values are significantly lower in the SFRs along the molecular ridge. The \oiiicii\ ratio is on average 3.2 for N158 and 2.3 for N160 and 2.2 for N159 - consistent with these regions being less energetic than \xdor. The three bright SFRs exhibit a wide range of \oiiicii\ ratios. In some areas, the ratio is particularly low because  the \oiii\ emission declines sharply beyond the H II regions, where the ionizing radiation is less energetic and conditions are less favorable for producing \oiii\ emission, due to its relatively high ionization potential of $\sim$35 eV.\\
The remaining points, corresponding to more diffuse emission (yellow, brown, and gray), generally exhibit higher \oiiicii\ ratios, with a trend of increasing ratios at lower \tir\ values. Due to the limited area where \oiii\ has been detected, this does not show a general trend but rather the existence of areas in the diffuse parts of the molecular ridge with relatively high \oiiicii\ compared to the bright star formation regions. The \oiii\ emission of these areas is, on average, 4.1 times  stronger than the \cii\ emission. Lower ratios are expected in the regions without clear detections as shown by the ratios based on the upper limits for \oiii. Similar to \xdor\ \citep{chevance20}, the existence of relatively strong \oiii\ detections compared to \cii\ in the extended, diffuse regions of the molecular ridge, indicates the high porosity of the low-Z environment, where high energy photons can travel significant distances before interacting with the ISM. The \oiiicii\ ratio is higher in the more diffuse regions, where \ciitir\ is elevated as well, also suggesting the presence of a diffuse UV field permeating a porous, clumpy ISM. This is further supported by the small-scale clumpy structure observed in the CO (2-1) map (Section~\ref{co_cii}). This conclusion is in agreement with those of \cite{Poglitsch95} and \cite{chevance16} for \xdor\ and \cite{lebouteiller12} for N11B.

\subsection{\cii\ compared to CO}
\label{co_cii}
The \ciico\ ratio has been used as a tracer of global star formation activity in galaxies \citep[e.g.,][]{stacey91}.  In normal star-forming galaxies this ratio typically ranges from  $\sim$ 1000 to 4000, with higher values observed in more active starburst galaxies (e.g. \citealt{stacey91}; \citealt{Stacey2010}; \citealt{Hailey2010}). In contrast, dwarf galaxies show ratios as high as 80000 \citep{madden20}.\\ 
Figure \ref{fig:ratio_diagrams} right shows the \cii\ luminosity versus the CO luminosity (where both \cii\ and CO (2-1) have been detected) normalized by \ltir. To estimate the CO (1-0) luminosity we used the CO (2-1) ALMA data from Tarantino et al. (in preparation) and Chen et al. (in preparation) and assumed a CO (2-1) / CO (1-0) ratio of 0.6 in K/(km/s) (e.g. \citealt{denBrok2021}, \citealt{Maeda2022}). \cii\ is much more extended and diffuse than CO, which leads to a large range of \ciico\ across the map, from 300 to over 100,000. While both \cii\ and CO peak locally on the \ltir\ peaks, CO exhibits filamentary structures not associated with either \cii\ or \ltir\ emission and \cii\ is present where no CO exists, as well. Some of these regions, framed in brown in Fig. \ref{fig:rgb}, show large variations of \cotir\ and \ciico\ (300 - 25000) ratio as well as the highest \cotir\ and the lowest \ciico\ values. Even within the three bright SFRs, the \ciico\ ratios vary significantly, from about 1000 up to over 100000. The average ratios of the SFRs, N158 (17000), N160 (12000) and N159 (8000) closely match the overall molecular ridge average value of 11000. When computed pixel by pixel, the average ratio is even higher (25500), with a tendency towards higher ratios in the more diffuse areas. 

\begin{figure*}
\centering
\includegraphics[width=0.48\textwidth]{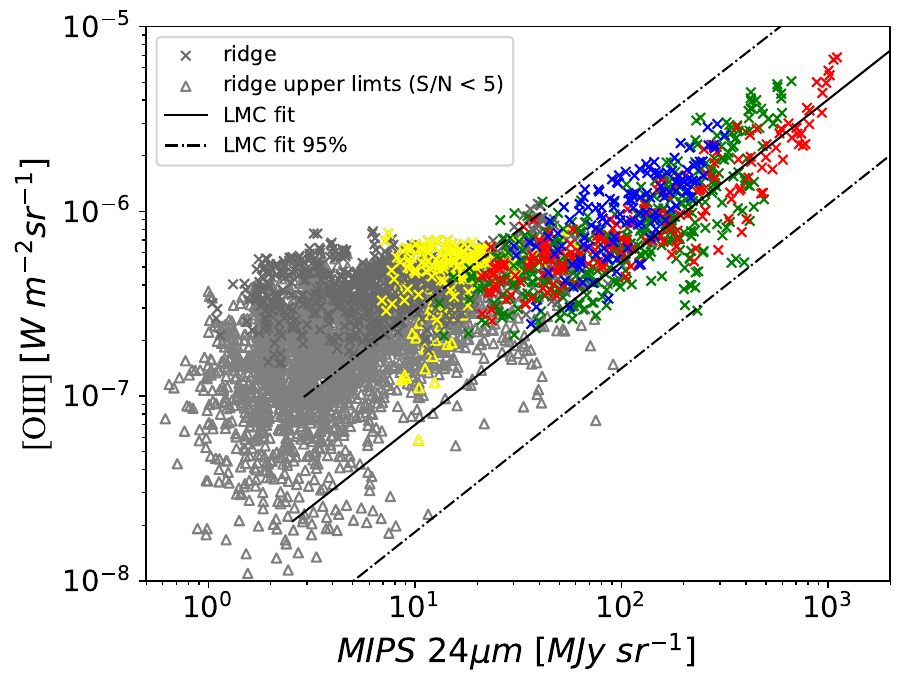}
\includegraphics[width=0.48\textwidth]{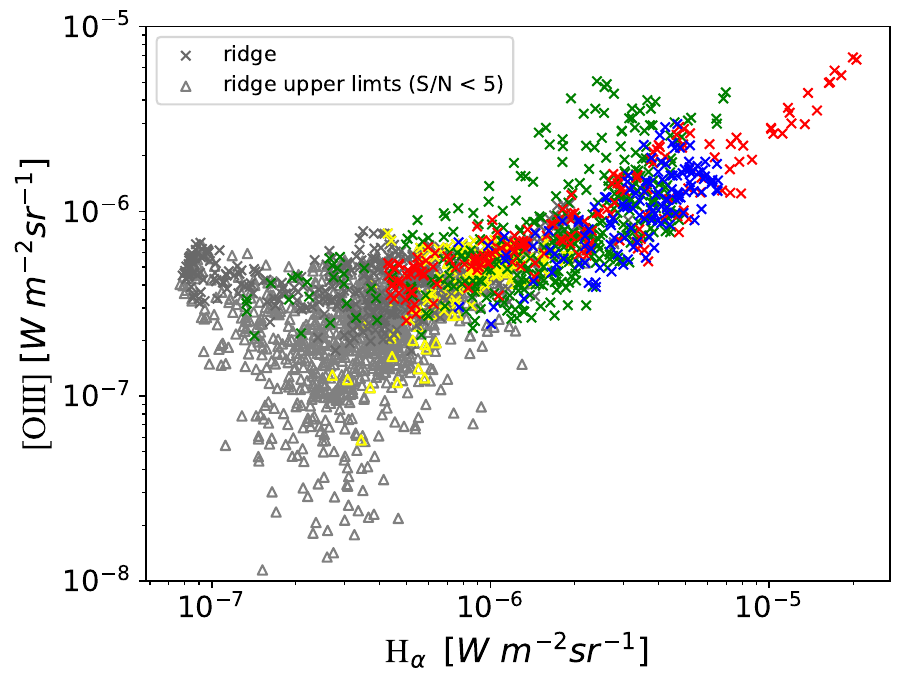}
\caption{\oiii\ emission versus 24\mic\  continuum emission (left) and versus H${\alpha}$ (right) in the three star formation regions in 12\arcsec\ × 12\arcsec\ pixels and at 48\arcsec\  resolution (in 24\arcsec\ × 24\arcsec\ pixels) for the rest of molecular ridge map with a detection more than 5$\sigma$ in both. Yellow, blue, red and green colors refer to the regions outlined in Figs. \ref{fig:rgb} and \ref{fig:LTIR}. Gray points represent the pixels in the ridge outside of these identified regions. For the diffuse regions shown in gray and yellow, fit results with a S/N below 5 are shown as upper limits. The black lines in the left plot show the fitted linear relation and the  95\% confidence interval from \cite{lamberthuyghe22}.}
\label{fig:oiii_24}
\end{figure*}

While the \ciico\ ratios in the molecular ridge (300 to 100,000) are lower compared to those found in \xdor\ (50,000-200,0000 \citep{chevance16}), they are very high compared to normal star-forming galaxies(1000-4000). Both the molecular ridge overall, as well as the three SFRs, exhibit values of \ciico\ ( as well as \ciitir\ and \cotir) typical of the ratios integrated over the low-Z galaxies from the DGS sample \citep{cormier15, madden20, ramambason24}. The extreme range of \ciico\ ratios in the molecular ridge reflects significant variations in CO emission. The \cotir\ ratios span more than 2 orders of magnitude, the spread being driven, for the most part, by variations of CO, rather than variations in \ltir. Across the map, \ciico\ values vary by 3 orders of magnitude, with higher ratios in regions with low \lco, holding important implications for CO-dark gas as well as the multiphase nature of \cii. This will be explored further in an upcoming paper.

\subsection{\oiii\ compared to 24$\mu m$ continuum emission}
\cite{lamberthuyghe22} demonstrated a correlation between \oiii\ emission and 24\mic\ continuum emission using spatially-resolved \hers/PACS \oiii\ maps of SFRs throughout the LMC. While \cite{lamberthuyghe22} include N158, N159, and N160 in their study, the \oiii\ observations were limited strictly to the  bright star formation sites, not going beyond. Their study also included the unresolved galaxy-wide low-Z DGS galaxies. 
While \oiii\ traces the ionized gas, the $\sim$24\mic\ continuum emission is emitted, for the most part, by the warm dust and is considered to be one of the most reliable tracers of the star formation rate in galaxies (e.g., \cite{kennicutt09}, \cite{Whitcomb2023}, \cite{Belfiore2023}). With the new data presented in this paper, we extend the analysis of \cite{lamberthuyghe22} to the surrounding medium beyond the previously targeted SFRs only, investigating the behavior of the 24\mic\ warm dust and the \oiii\ in the more extended,  diffuse, low-Z environments.\\
Figure~\ref{fig:oiii_24} shows that the \oiii\ line emission and the 24\mic\ continuum emission correlate well within the three bright SFRs, beyond the small areas of peak \oiii\ emission mapped with \hers/PACS, and, in general, follow the correlation shown by  \cite{lamberthuyghe22} and also seen in some LMC regions detected by \akari\ \citep{Kawada11}\footnote{They compare \oiii\ to the 88\mic\ continuum, which they demonstrate exhibits a strong correlation with the 24\mic\ continuum.}. However, in the more extended, diffuse regions (gray and yellow markers), where the 24\mic\ emission has fallen off and \oiii\ remains well-detected, we find a deviation from the rather tight correlation of \oiii\ and 24\mic\ toward the bright star formation sites. The \oiii\ $-$ 24\mic\ relation has a tendency to flatten further from the star-formation sites. \oiii\ emission can exist, where the hot dust emission arising from 24\mic\ does not reach. The dust and the highly excited \oiii\ seem to be well-mixed around the SFRs, but less so outside these regions.\\
Related to this, we do see a correlation between \oiii\ and \halpha\ (Fig. \ref{fig:oiii_24}) for the brighter SFRs as well. For lower values of \halpha, \oiii\ is also relatively more prominent, pointing to a flattening out of the correlation seen for higher values of \halpha,  similar to \oiii\ vs 24\mic. Overall, the individual SFRs and the extended regions have different behaviors in \oiii\ vs 24\mic\ and \oiii\ vs \halpha\ (and \oiii\ vs \cii; see section \ref{oiii_cii}). This will be the subject of a followup study to investigate the roles of the excitation sources/radiation field, the distribution of dust and gas densities, and porosity in the observations. 
 
\section{Conclusions}\label{conclusions}
The SOFIA Legacy Project, \lmcplus, has mapped a 610pc $\times$ 260pc region in the molecular ridge in the LMC in \cii\ and \oiii\ at 2.5 pc resolution, using FIFI-LS. This paper presents the first results from these observations, including the line flux maps, details of the observational strategy and the data reduction procedures, for which some new techniques have been developed. Both the \ciiline\ and \oiiilineup\ lines are detected widespread throughout the region. This first analysis compares the \lmcplus\ observations with maps of other tracers, including  CO (2-1) from ALMA (Tarantino et al. in preparation; Chen et al. in preparation), \spitz\ 24 \mic, and \ltir\ (Belloir et al. in preparation). We summarize our findings here:
 
\begin{itemize}
    \item While the \cii\ emission peaks toward the three prominent star-forming sites, N158, N159, and N160, it is detected well beyond these concentrations, throughout the \lmcplus\ map.  The importance of \cii\ as a cooling line relative to \ltir\ increases with decreasing \ltir. Across a wide variety of environments and different size scales, we see a consistent, tight correlation between \ciitir\ and \ltir\ over three orders of magnitude of \ltir,  with increasingly high values of \ciitir\ for lower values of \ltir. This trend points to a strong link between the "\cii-deficit" and the  local physical conditions instead of global properties.
    \item We find extended bright \oiii\ emission throughout the region, even beyond the three bright star formation regions. Where \oiii\ is detected, it is almost always brighter than \cii. The \oiiicii\ ratio reaches 10 in the extended, more diffuse regions, indicating the presence of a diffuse UV field within a porous and clumpy ISM. However, the ridge values of \oiiicii\ do not reach the extreme ratios of $\sim$100 found in \xdor\ \citep{chevance20}.  
    \item We find the \ciico\ ratio extremely high in most of the molecular ridge as well as displaying a large variation of over three orders of magnitude with high \ciico\ preferring regions with low \lco, possibly holding important implications for CO-dark gas.
    \item We confirm the correlation between \oiii\ and 24\mic\ found by \cite{lamberthuyghe22} who targeted only the bright SFRs. Where we detect \oiii\ emission outside of the brightest regions, it is not correlated with 24\mic, but remains high as the 24\mic\ emission decreases, another indication of the high porosity of the ISM in this environment.
    
\end{itemize}
This paper presents the first results of the \lmcplus\ project, with the aim of characterizing the physical conditions and thermal processes across the different phases of the ISM. By zooming into our nearest low-Z galaxy, an ideal laboratory for such studies, we can understand how local ISM properties influence star formation on parsec scales. Future papers will provide detailed modeling and analyses to further explore these goals.

\section{Data availibility}
Line flux maps of \cii\ and \oiii\ are available in electronic form at the CDS via anonymous ftp to \url{cdsarc.u-strasbg.fr} (130.79.128.5) or via \url{http://cdsweb.u-strasbg.fr/cgi-bin/qcat?J/A+A/}.

\begin{acknowledgements}
We acknowledge the dedication of the whole SOFIA 2022 Chile deployment team and the local staff in Santiago de Chile to making the SOFIA observations happen under the challenging boundary conditions of the flight series. We also thank the anonymous referee for the useful and constructive comments and suggestions, which helped improve the clarity and quality of the manuscript.\\SOFIA, the "Stratospheric Observatory for Infrared Astronomy" is a joint project of the Deutsches Zentrum für Luft- und Raumfahrt e.V. (DLR, German Aerospace Centre; grants 50OK0901, 50OK1301 and 50OK1701) and the National Aeronautics and Space Administration (NASA). This paper makes use of the following ALMA data: ADS/JAO.ALMA\#2018.A.00061.S. ALMA is a partnership of ESO (representing its member states), NSF (USA) and NINS (Japan), together with NRC (Canada), NSTC and ASIAA (Taiwan), and KASI (Republic of Korea), in cooperation with the Republic of Chile. The Joint ALMA Observatory is operated by ESO, AUI/NRAO and NAOJ. It is funded on behalf of DLR by the Federal Ministry for Economic Affairs and Energy based on legislation by the German Parliament and funded by the state of Baden-Württemberg and the Universität Stuttgart. Scientific operation for Germany is coordinated by the German SOFIA Institute (DSI) of the Universität Stuttgart, in the USA by the Universities Space Research Association (USRA). Part of this work has been funded by the NASA grant number 09\_0036. MC and LR gratefully acknowledge funding from the DFG through an Emmy Noether Research Group (grant number CH2137/1-1). COOL Research DAO \citep{cool_whitepaper} is a Decentralized Autonomous Organization supporting research in astrophysics. MR wishes to acknowledge partial support from ANID(CHILE) through Basal FB210003. T.W. acknowledges financial support from the University of Illinois Vermilion River Fund for Astronomical Research. FG acknowledges support by the French National Research Agency under the contracts WIDENING (ANR-23-ESDIR-0004) and REDEEMING (ANR-24-CE31-2530), as well as by the Actions Thématiques ``Physique et Chimie du Milieu Interstellaire'' (PCMI) of CNRS/INSU, with INC and INP, and ``Cosmologie et Galaxies'' (ATCG) of CNRS/INSU, with INP and IN2P3, both programs being co-funded by CEA and CNES.
\end{acknowledgements}

\bibliographystyle{aa} 
\bibliography{lmc} 

\appendix

\section{Sample spectra from the \cii\ and \oiii\ maps} \label{spectra}
In this section we show sample spectra for both \cii\ and \oiii\ at a total of seven locations throughout the map. The locations are marked and labeled in both plots of Fig. \ref{fig:fifi_maps} and represent the dynamic range of fluxes detected in both lines. For \cii\ and \oiii\ in the three bright star formation regions, fluxes are extracted and fitted in 15.3\arcsec apertures. For \oiii\ at the four locations in the diffuse parts the fluxes are extracted and fitted in 48\arcsec apertures. The fit results are shown by the black dashed lines. For \oiii\ in addition to the overall fit result the baseline component is shown by a red line to aid the visual separation of the astronomical line from the baseline. Details on the line fitting are described in chapter \ref{line_fitting}. 
\begin{figure}[H]

\centering
\includegraphics[width=0.4\textwidth]{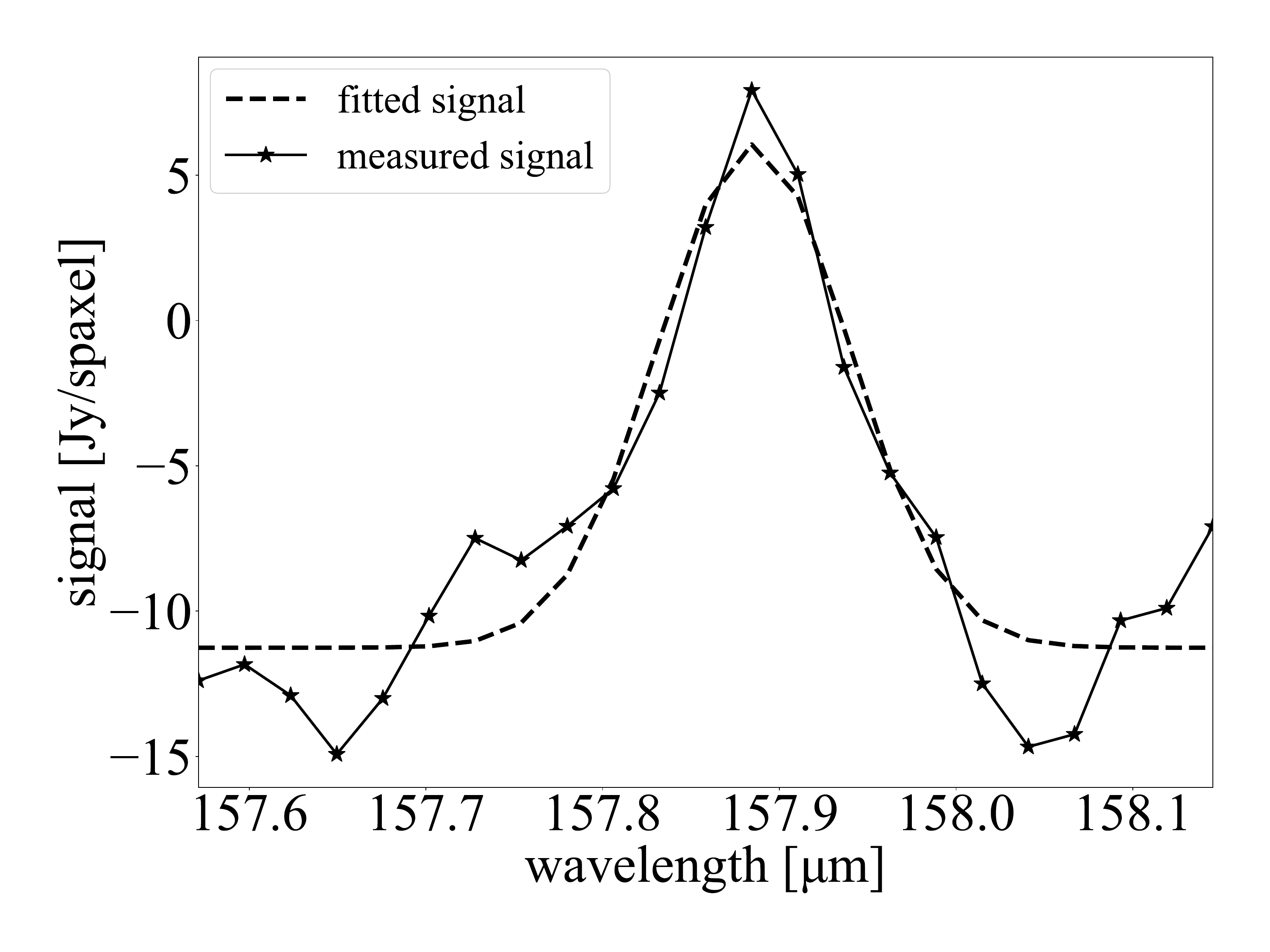}
\includegraphics[width=0.4\textwidth]{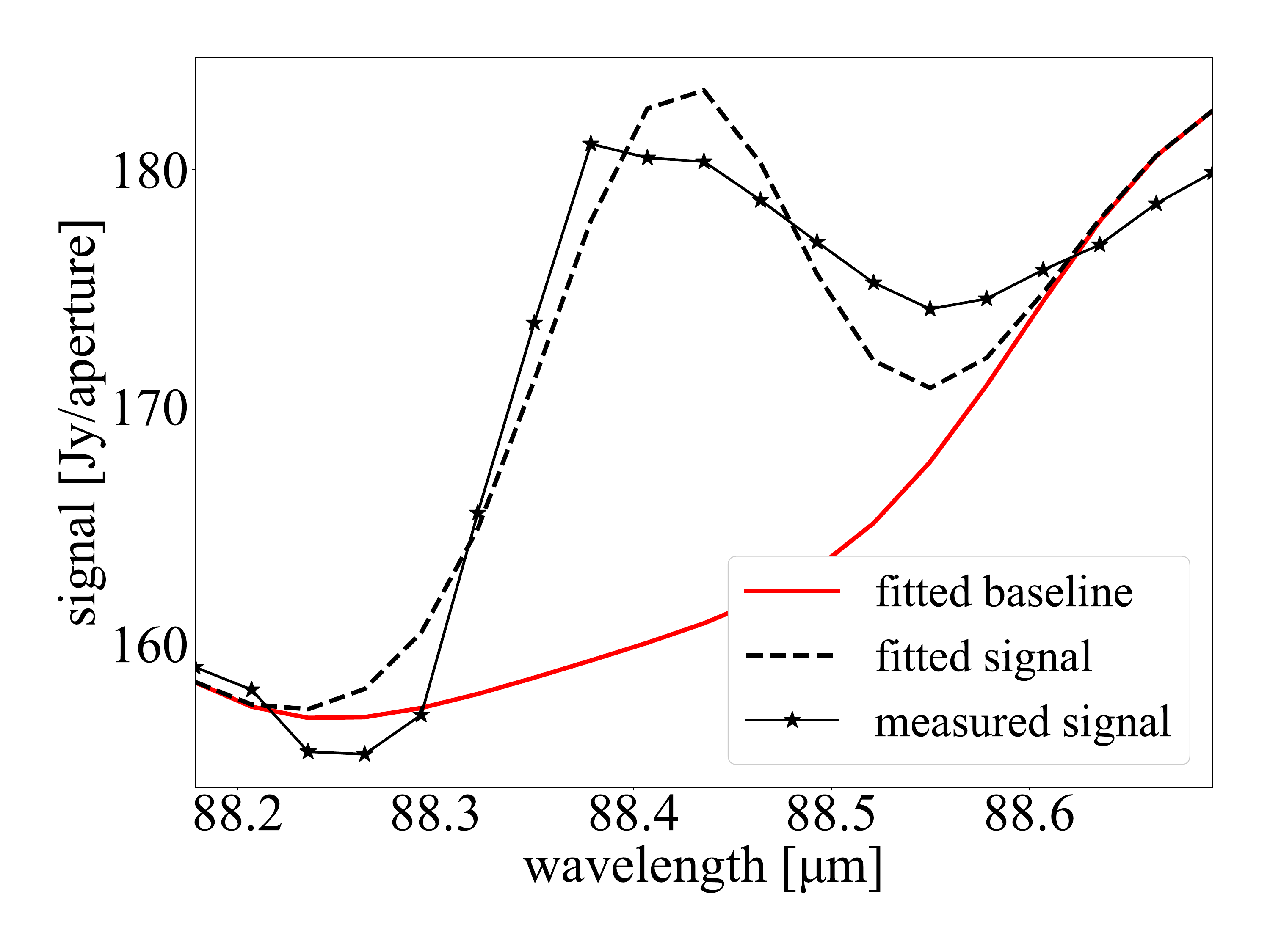}

\caption{
Spectra from location N158 indicated by the crosses in Fig. \ref{fig:fifi_maps} averaged in apertures drawn around the location with a 15\arcsec\ diameter. The measured spectra are shown by the connected black stars. The dashed black line shows the fitted signal. For the \oiii\ spectra the fitted baseline is additionally shown as a red line. Details on the line fit can be found in section \ref{line_fitting}. {\it Top: } \cii\ {\it Bottom: } \oiii
}
\label{fig:spectra11}

\end{figure}

\begin{figure}

\centering
\includegraphics[width=0.4\textwidth]{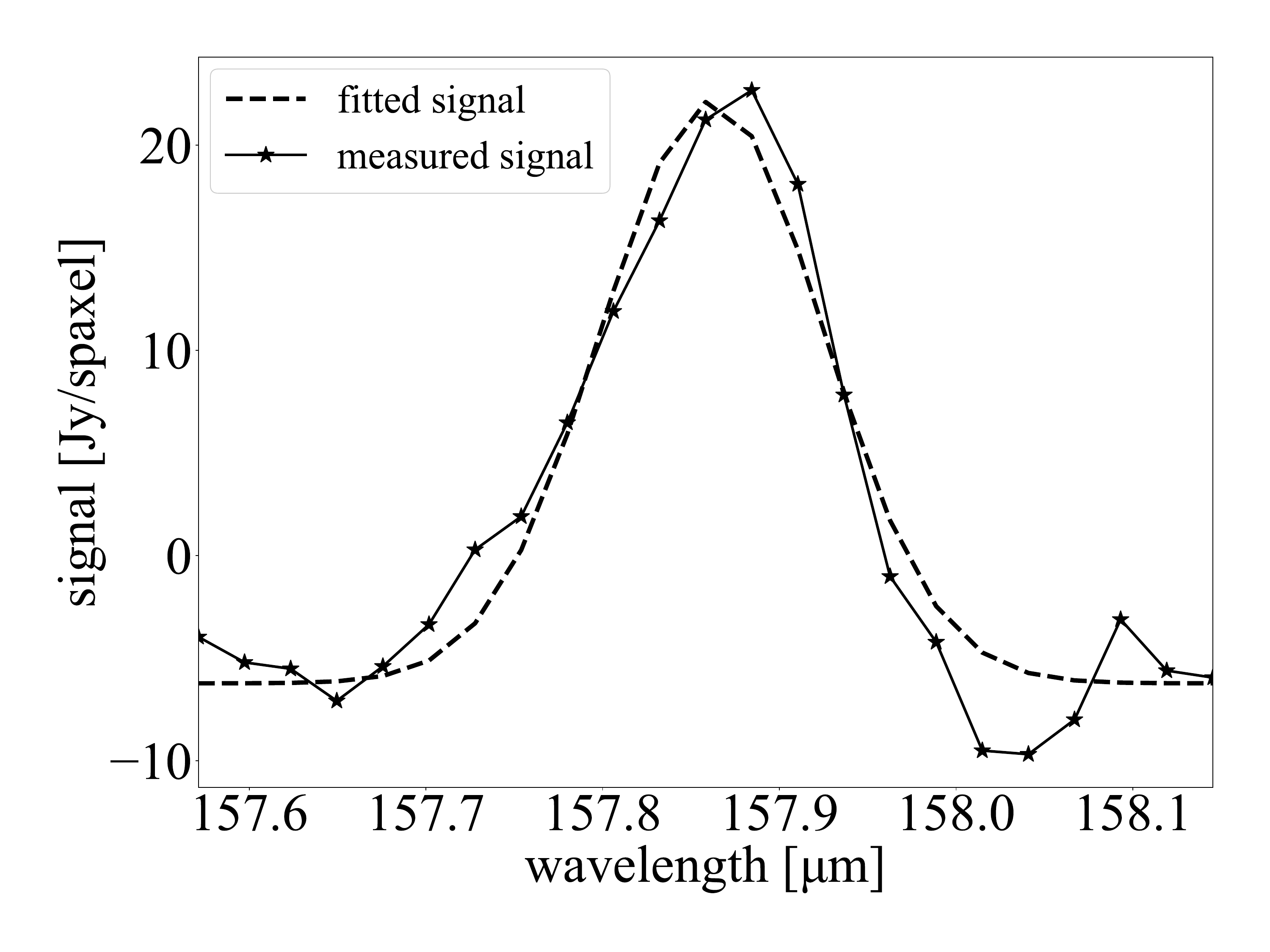}
\includegraphics[width=0.4\textwidth]{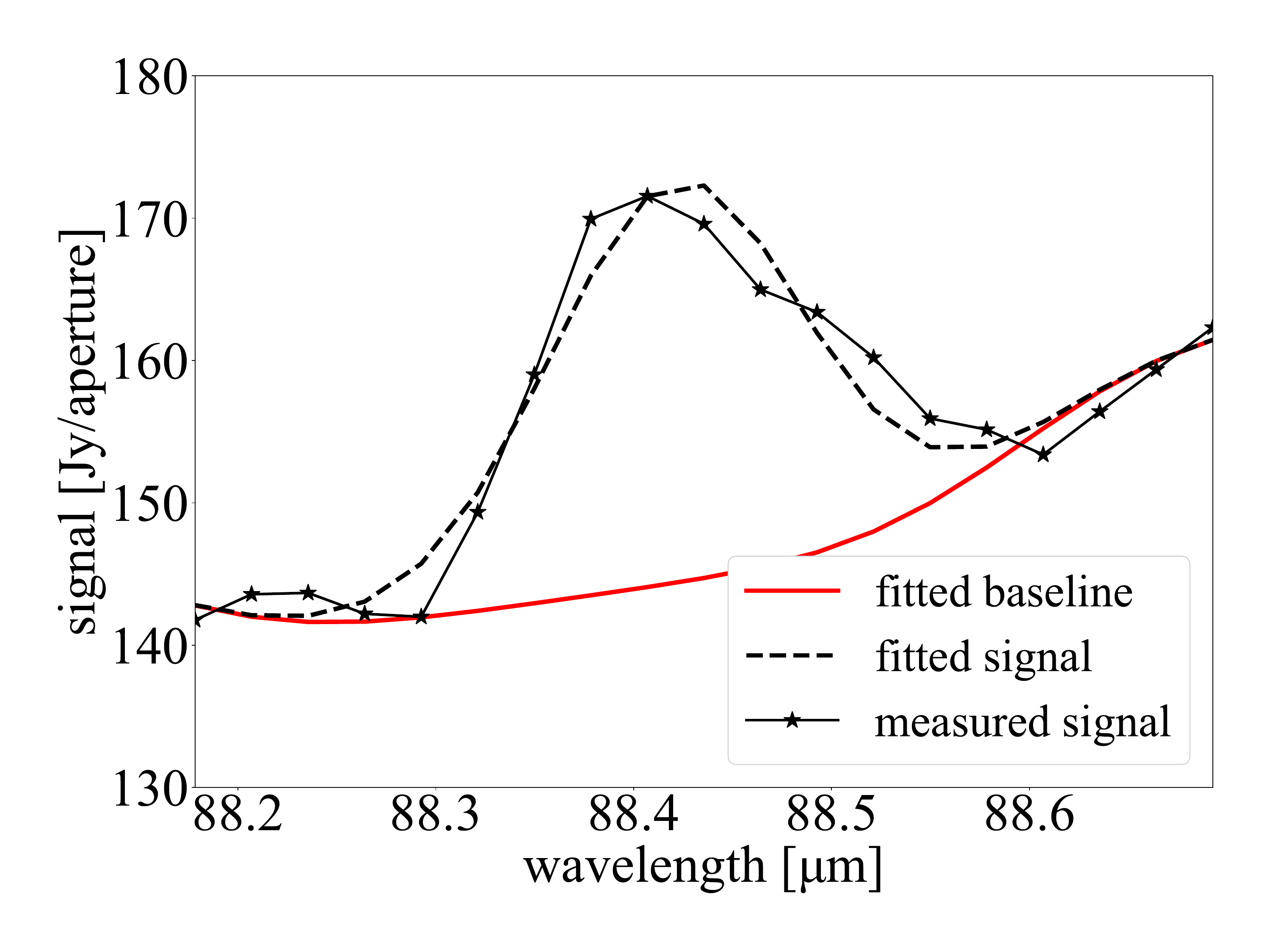}
\includegraphics[width=0.4\textwidth]{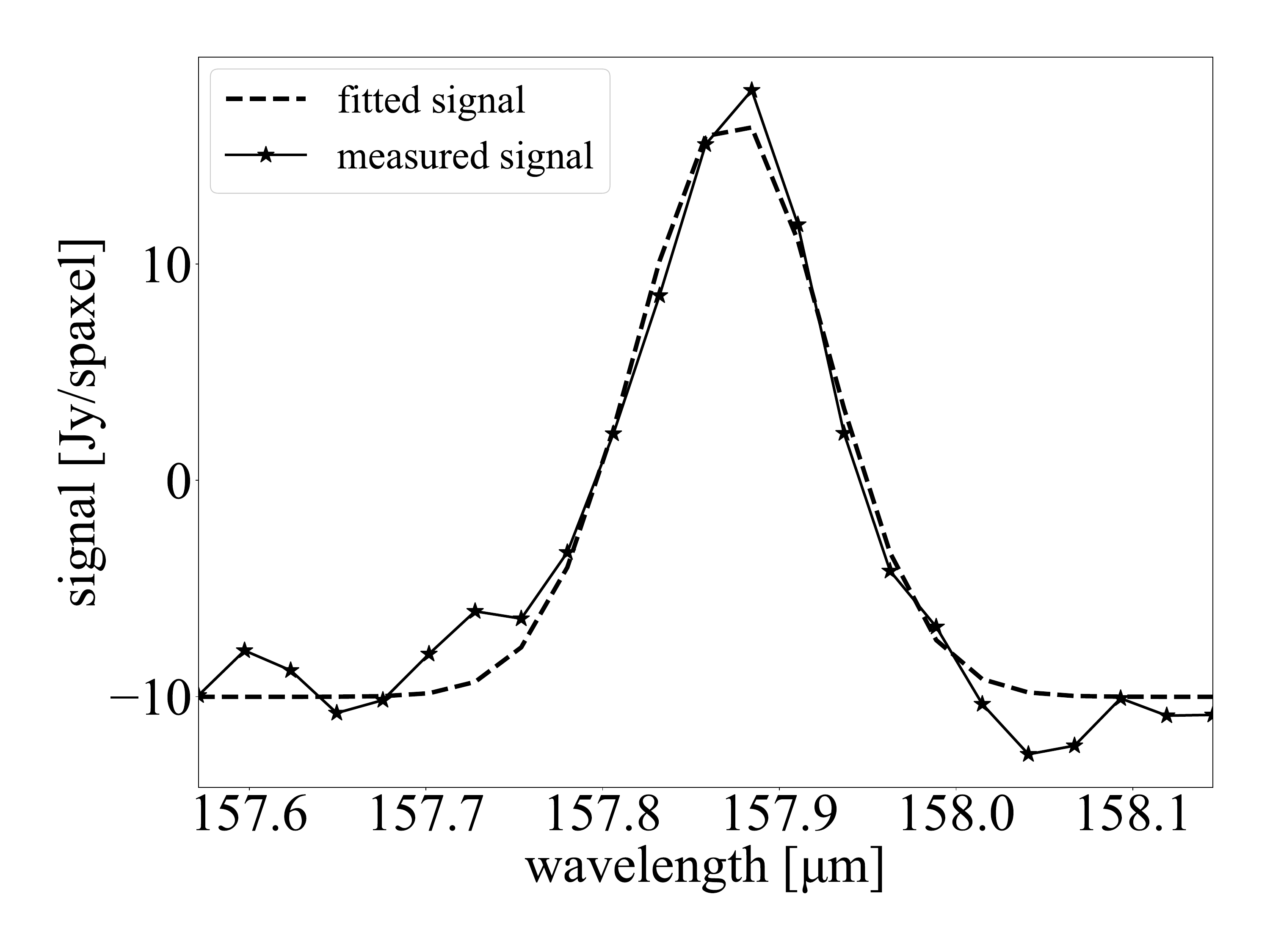}
\includegraphics[width=0.4\textwidth]{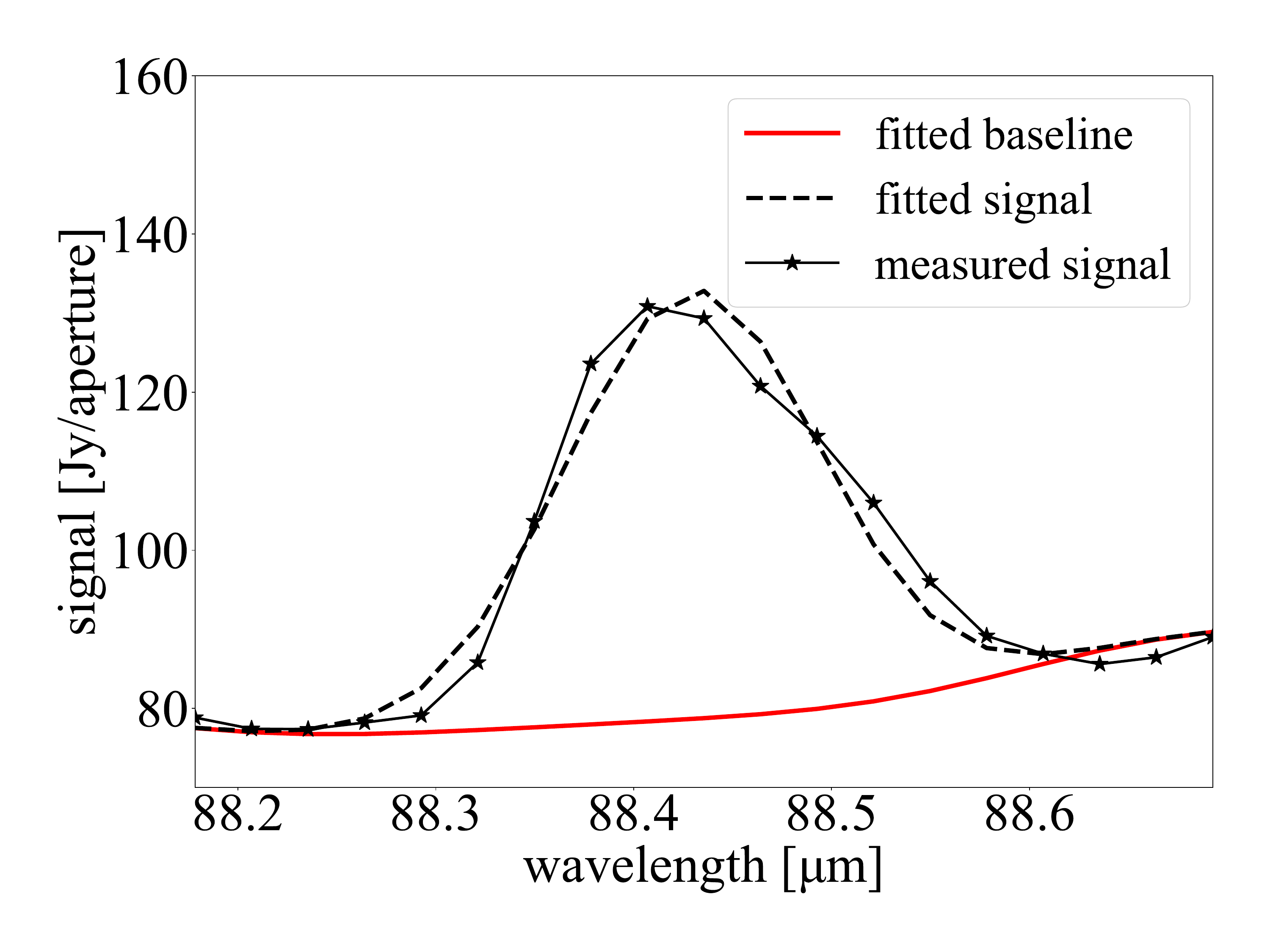}

\caption{
same as figure \ref{fig:spectra11};
{\it Top 2: }N160; {\it Bottom 2:} N159}

\label{fig:spectra12}

\end{figure}

\begin{figure*}
\centering
\includegraphics[width=0.40\textwidth]{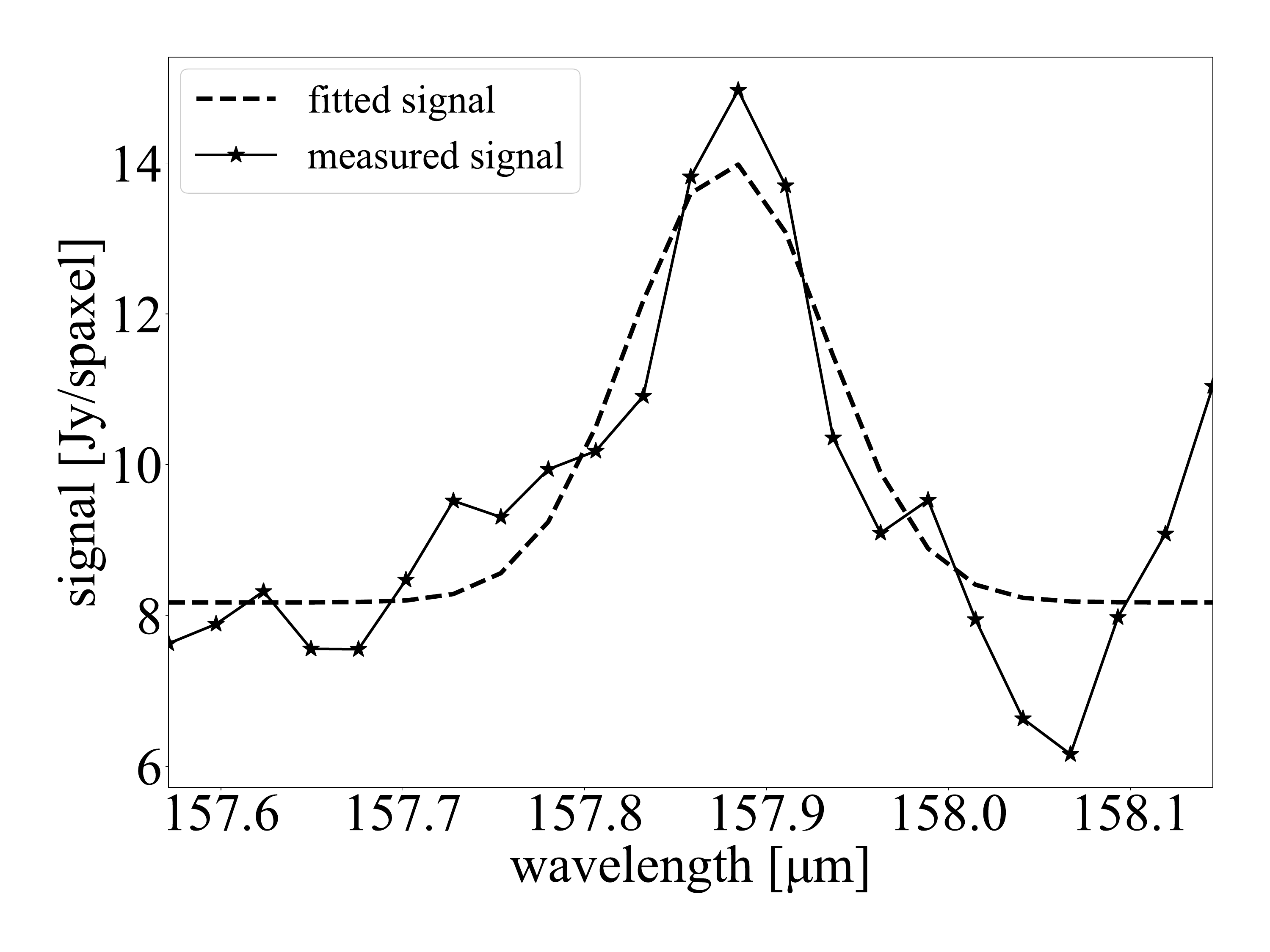}
\includegraphics[width=0.40\textwidth]{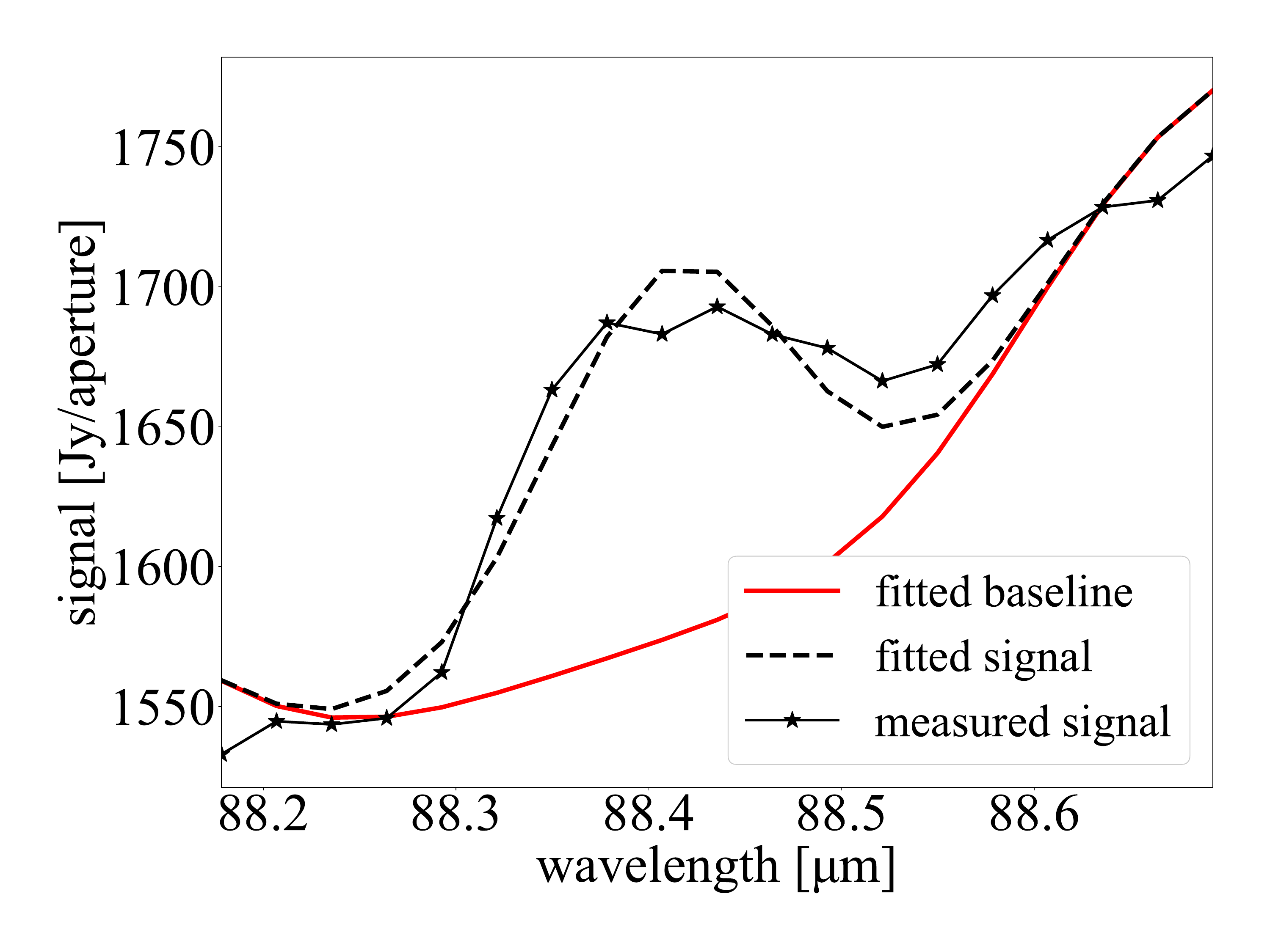}
\includegraphics[width=0.40\textwidth]{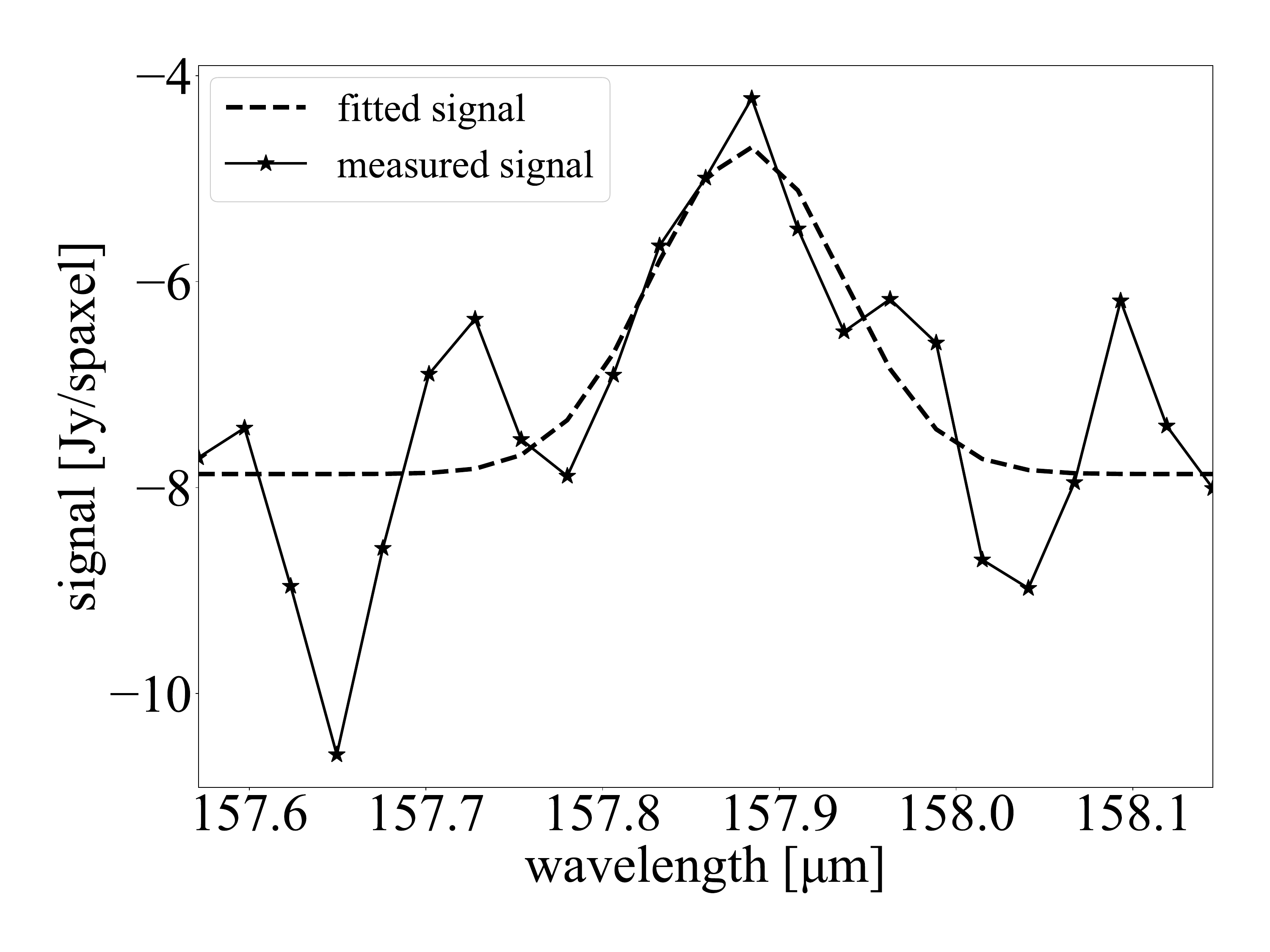}
\includegraphics[width=0.40\textwidth]{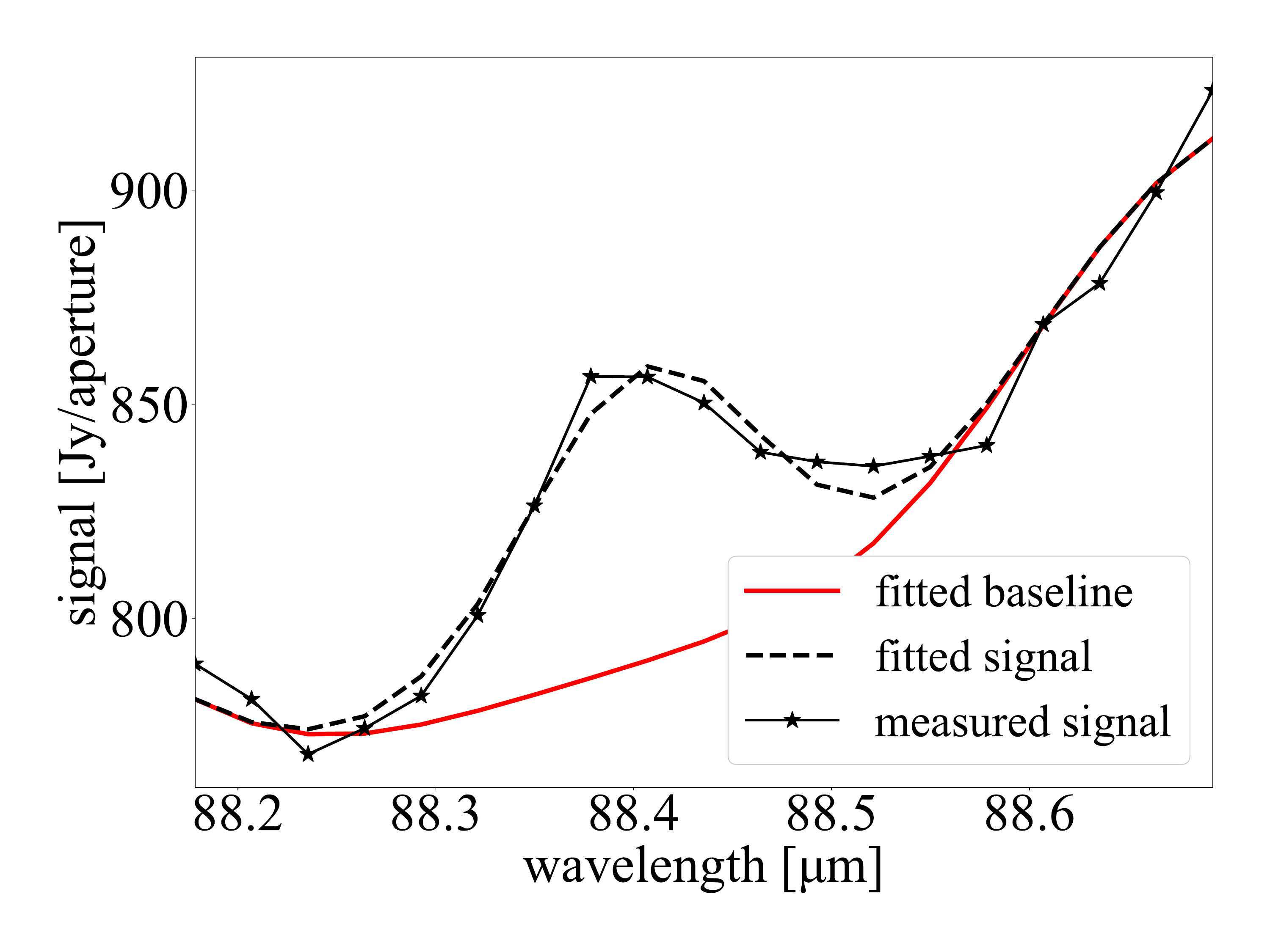}
\includegraphics[width=0.40\textwidth]{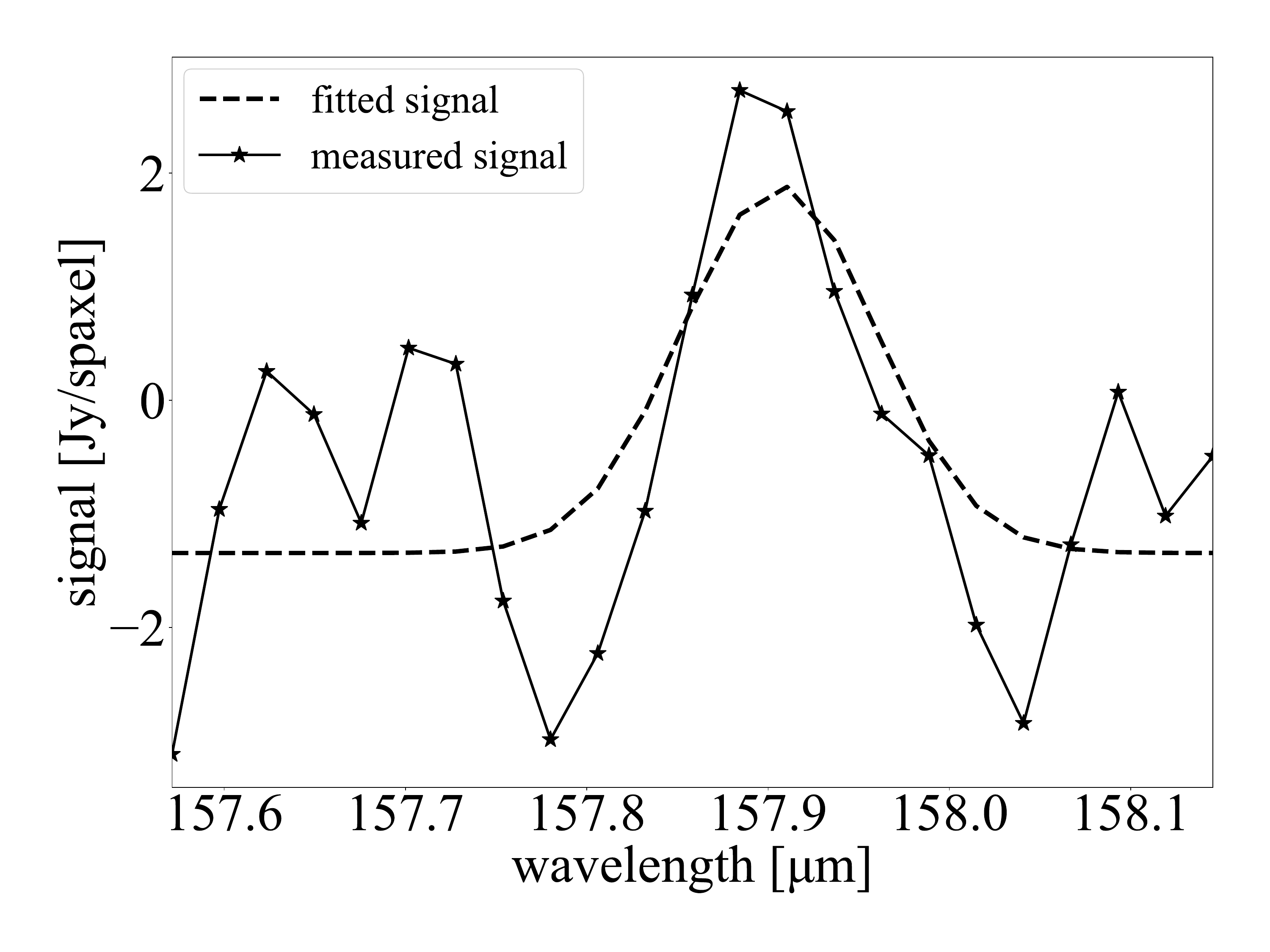}
\includegraphics[width=0.40\textwidth]{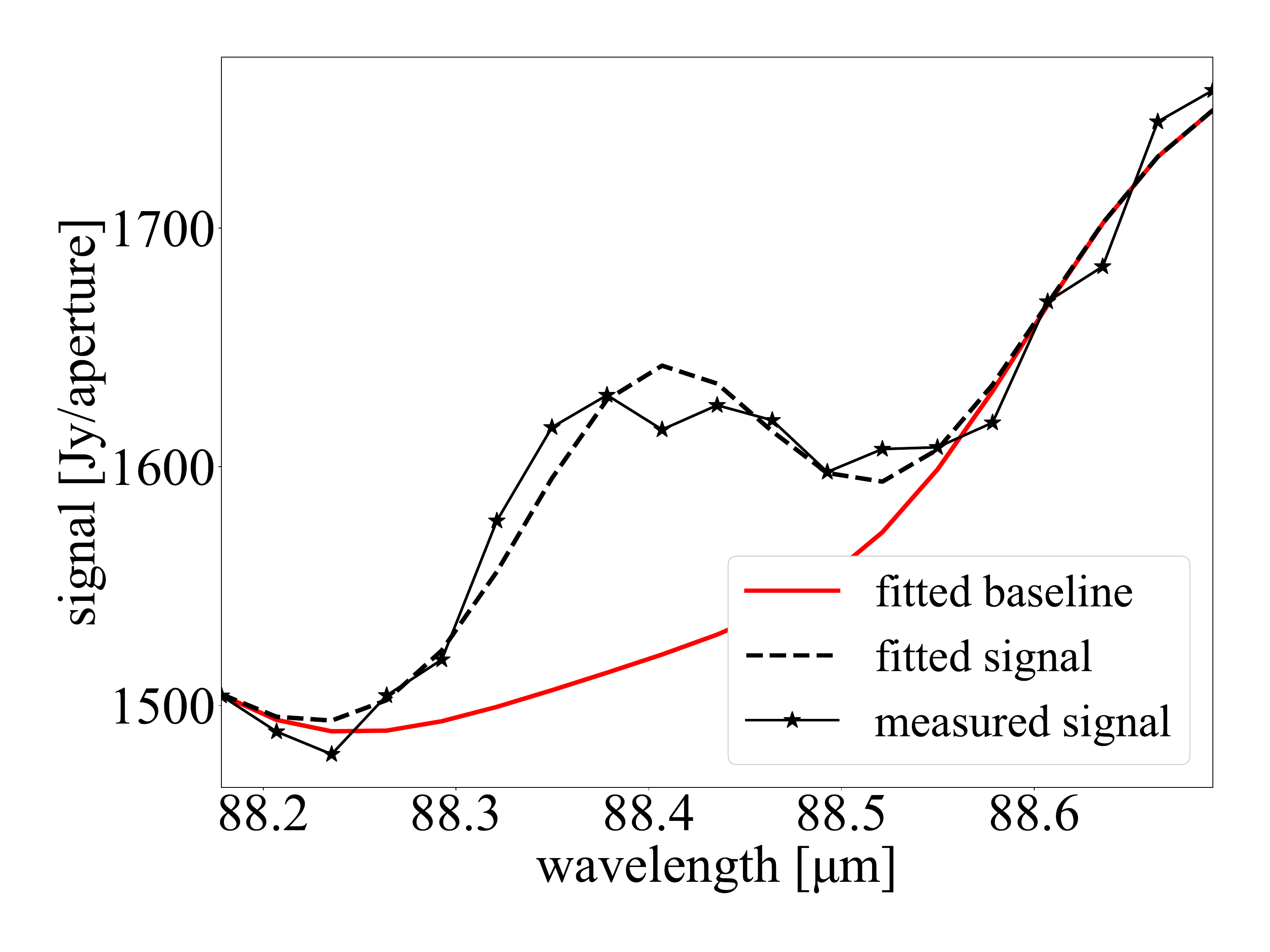}
\includegraphics[width=0.40\textwidth]{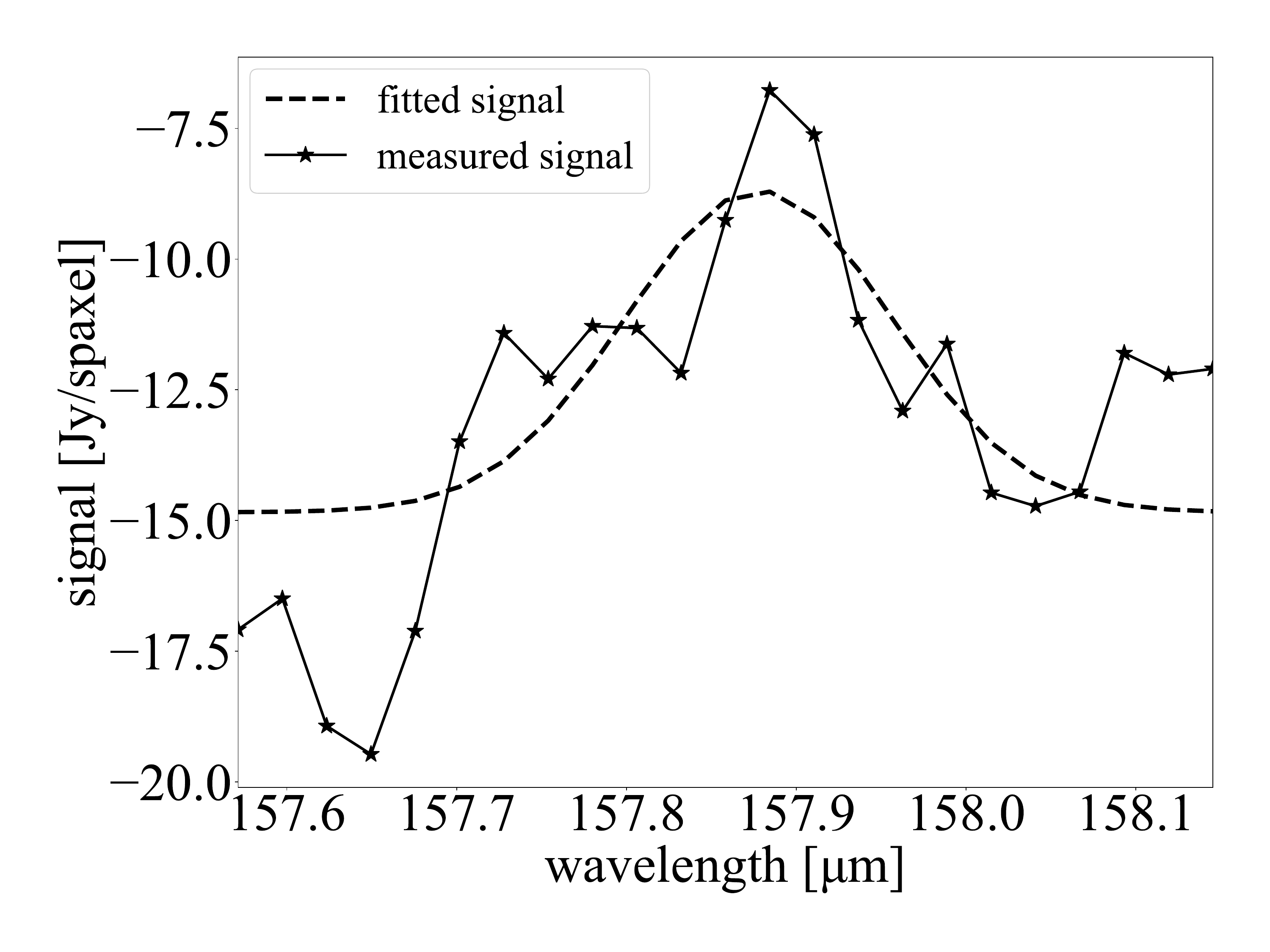}
\includegraphics[width=0.40\textwidth]{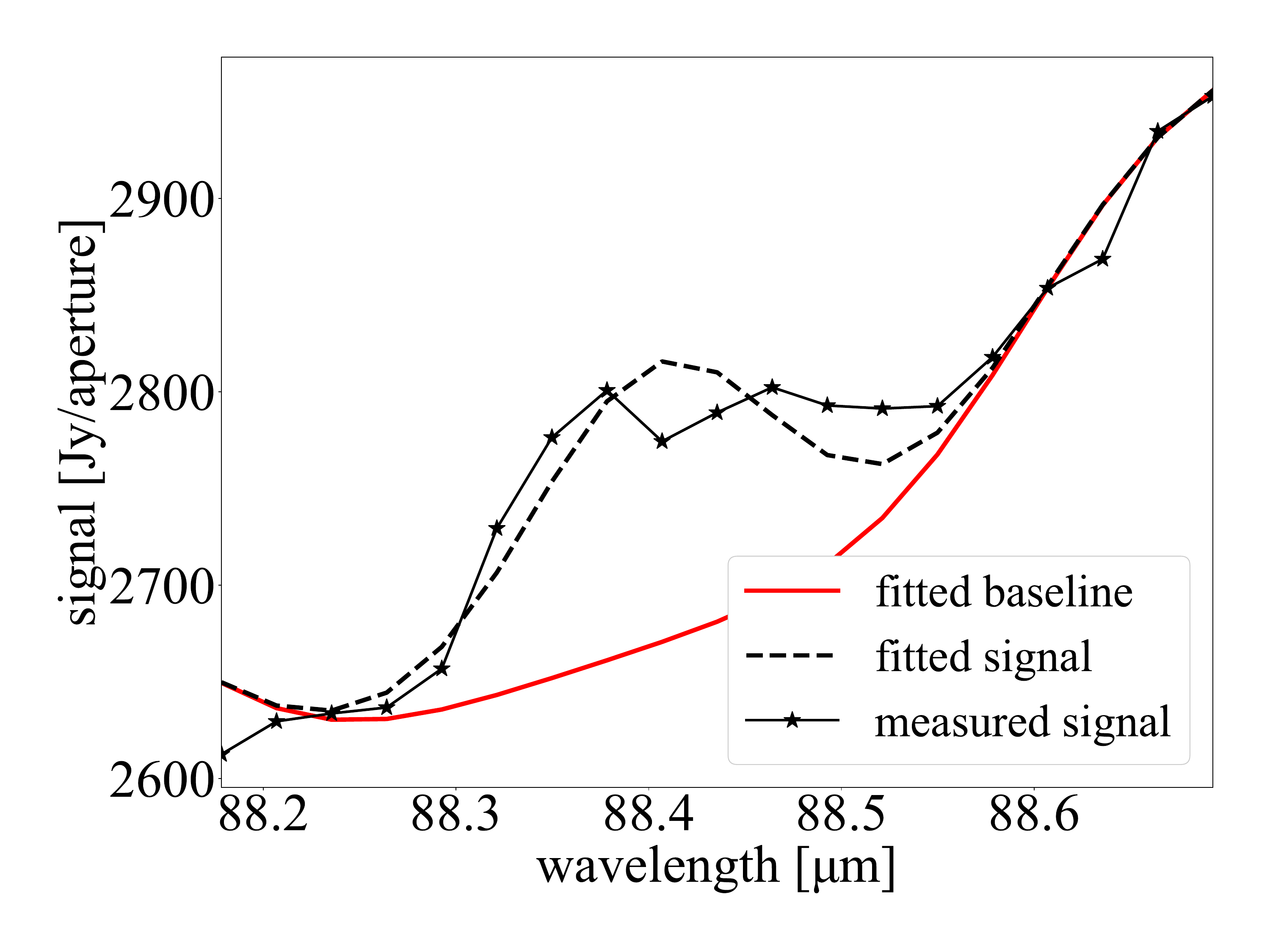}
\caption{
Same as figure \ref{fig:spectra11} but with aperture diameters of 48\arcsec\ for \oiii. 
{\it from top to bottom: } dif1, dif2, dif3, dif4
}

\label{fig:spectra2}
\end{figure*}

\clearpage

\section{Cross calibration with \textit{Herschel}/PACS}

This appendix reports the flux measurements performed in 20 different apertures observed by both \sofia/FIFI-LS and \hers/PACS. 

\begin{table}[H]
\caption{PACS vs FIFI-LS cross-correlation\label{table:PACS_cross}} 
\centering 
\begin{tabular}{c c c c c c c}
\hline \hline 
line & R.A. & Dec. & radius & line flux FIFI-LS & line flux PACS  & $(F_{FIFI}-F_{PACS})/F_{PACS}$\\
~ & (J2000)& (J2000) & (arcsec) & ($10^{-14}$W~m$^{-2}$) & ($10^{-14}$W~m$^{-2}$) &$(\%)$\\ 
\hline
~ \oiii & 05h39m13.46s & -69d30m37.84s & 18 &  5.5$\pm$0.6 & 5.21$\pm$0.02 & 5.6 \\
~ \oiii & 05h39m13.62s & -69d29m52.35s & 18 &  3.0$\pm$0.5 & 3.05$\pm$0.04 & -1.6\\
~ \oiii & 05h39m44.40s & -69d38m42.68s & 18 &  16.2$\pm$1.1 & 16.29$\pm$0.09 & -0.6 \\
~ \oiii & 05h39m35.57s & -69d39m17.57s & 18 &  5.1$\pm$0.8 & 4.93$\pm$0.04 & 3.4 \\
~ \oiii & 05h39m37.90s & -69d45m23.00s & 18 & 1.8$\pm$0.4  & 1.90$\pm$0.01 & -5.2 \\
~ \oiii & 05h39m39.98s & -69d46m21.13s & 18 &  7.0$\pm$0.8  & 7.29$\pm$0.04 & -4.0 \\
~ \oiii & 05h39m53.61s & -69d45m19.57s & 18 &  7.6$\pm$0.6  & 6.97$\pm$0.04 & 9.0 \\
~ \oiii & 05h40m07.77s & -69d44m58.83s & 18 &  9.6$\pm$0.6  & 8.96$\pm$0.04 & 7.1 \\
\hline
~ &  &  &  &  & mean deviation: & 4.6$\%$\\
\hline
~ \cii & 05h39m16.31s & -69d30m39.67s & 18 & 1.8$\pm$0.2 & 1.78$\pm$0.02 & 1.1\\
~ \cii & 05h39m15.49s & -69d29m58.11s & 18 & 1.53$\pm$0.04 & 1.53$\pm$0.02 & 0\\
~ \cii & 05h39m38.55s & -69d39m10.41s & 18 & 2.0$\pm$0.1 & 2.09$\pm$0.02& -4.3\\
~ \cii & 05h39m45.71s & -69d38m38.42s & 18 & 2.46$\pm$0.07 & 2.46$\pm$0.02 & 0\\
~ \cii & 05h39m37.96s & -69d46m11.78s & 18 & 2.02$\pm$0.09 & 2.08$\pm$0.02 & -2.9 \\
~ \cii & 05h39m37.26s & -69d45m27.63s & 18 & 1.62$\pm$0.03 & 1.66$\pm$0.02 & -2.4 \\
~ \cii & 05h39m56.75s & -69d45m29.81s & 18 & 1.37$\pm$0.03 & 1.36$\pm$0.01 & 0.7\\
~ \cii & 05h39m52.13s & -69d45m02.63s & 18 & 1.54$\pm$0.09 & 1.51$\pm$0.01 & 2.0\\
~ \cii & 05h39m46.52s & -69d44m42.04s & 18 & 1.7 $\pm$0.1 & 1.75$\pm$0.01 & -2.9\\
~ \cii & 05h40m04.85s & -69d44m37.70s & 18 & 2.18$\pm$0.04 & 2.18$\pm$0.02 & 0\\
~ \cii & 05h40m01.53s & -69d50m21.80s & 18 & 0.21$\pm$0.02 & 0.211$\pm$0.002 & -0.5\\	
~ \cii & 05h39m55.28s & -69d50m17.63s & 18 & 0.14$\pm$0.03 & 0.172$\pm$0.002 & -18.6\\
\hline 
~ &  &  &  &  & mean deviation: & 3$\%$\\
\hline 
\hline 
\end{tabular}
\end{table}
\end{document}